\documentclass[floatfix,twocolumn,aps,prl,reprint,superscriptaddress]{revtex4-1}
\usepackage{amsmath}
\usepackage{amssymb}
\usepackage{float}
\usepackage{graphicx}
\usepackage{color,soul}
\usepackage{hyperref}
\usepackage[colorinlistoftodos]{todonotes}
\usepackage{lipsum}
\usepackage{braket}

\graphicspath{ {figures/} {figures_supplement/} }

\makeatletter

\@ifundefined{textcolor}{}
{%
 \definecolor{BLACK}{gray}{0}
 \definecolor{WHITE}{gray}{1}
 \definecolor{RED}{rgb}{1,0,0}
 \definecolor{GREEN}{rgb}{0,1,0}
 \definecolor{BLUE}{rgb}{0,0,1}
 \definecolor{CYAN}{cmyk}{1,0,0,0}
 \definecolor{MAGENTA}{cmyk}{0,1,0,0}
 \definecolor{YELLOW}{cmyk}{0,0,1,0}
}
\usepackage{dcolumn}
\usepackage{bm}
\usepackage{float}

\usepackage{graphicx}
\makeatother

\begin{document}

\title{
Multimode photon blockade
}
\author{Srivatsan Chakram}
\thanks{These authors contributed equally to this work.}
\affiliation{James Franck Institute, University of Chicago, Chicago, Illinois 60637, USA}
\affiliation{Department of Physics, University of Chicago, Chicago, Illinois 60637, USA}
\affiliation{Department of Physics and Astronomy, Rutgers University, Piscataway, NJ 08854, USA}

\author{Kevin He}
\thanks{These authors contributed equally to this work.}
\affiliation{James Franck Institute, University of Chicago, Chicago, Illinois 60637, USA}
\affiliation{Department of Physics, University of Chicago, Chicago, Illinois 60637, USA}

\author{Akash V. Dixit}
\affiliation{James Franck Institute, University of Chicago, Chicago, Illinois 60637, USA}
\affiliation{Department of Physics, University of Chicago, Chicago, Illinois 60637, USA}
\affiliation{Kavli Institute for Cosmological Physics, University of Chicago, Chicago, Illinois 60637, USA}

\author{Andrew E. Oriani}
\affiliation{James Franck Institute, University of Chicago, Chicago, Illinois 60637, USA}
\affiliation{Pritzker School of Molecular Engineering, University of Chicago, Chicago, Illinois 60637, USA}

\author{Ravi K. Naik}
\affiliation{James Franck Institute, University of Chicago, Chicago, Illinois 60637, USA}
\affiliation{Department of Physics, University of Chicago, Chicago, Illinois 60637, USA}
\affiliation{Department of Physics, University of California Berkeley, California 94720, USA}

\author{Nelson Leung}
\affiliation{James Franck Institute, University of Chicago, Chicago, Illinois 60637, USA}
\affiliation{Department of Physics, University of Chicago, Chicago, Illinois 60637, USA}

\author{Hyeokshin Kwon}
\affiliation{Samsung Advanced Institute of Technology (SAIT), Samsung Electronics Co., Ltd, Suwon, Republic of Korea}

\author{Wen-Long Ma}
\affiliation{Pritzker School of Molecular Engineering, University of Chicago, Chicago, Illinois 60637, USA}

\author{Liang Jiang}
\affiliation{Pritzker School of Molecular Engineering, University of Chicago, Chicago, Illinois 60637, USA}

\author{David I. Schuster}
\email{David.Schuster@uchicago.edu}
\affiliation{James Franck Institute, University of Chicago, Chicago, Illinois 60637, USA}
\affiliation{Department of Physics, University of Chicago, Chicago, Illinois 60637, USA}
\affiliation{Pritzker School of Molecular Engineering, University of Chicago, Chicago, Illinois 60637, USA}

\begin{abstract}
Interactions are essential for the creation of correlated quantum many-body states. While two-body interactions underlie most natural phenomena, three- and four-body interactions are important for the physics of nuclei~\cite{loiseau1967three}, exotic few-body states in ultracold quantum gases~\cite{hammer2013colloquium}, the fractional quantum Hall effect~\cite{wojs2010global}, quantum error correction~\cite{kitaev2006anyons}, and holography~\cite{sachdev1993gapless,maldacena2016remarks}.
Recently, a number of artificial quantum systems have emerged as simulators for many-body physics, featuring the ability to engineer strong interactions. However, the interactions in these systems have largely been limited to the two-body paradigm, and require building up multi-body interactions by combining two-body forces. Here, we demonstrate a pure N-body interaction between microwave photons stored in an arbitrary number of electromagnetic modes of a multimode cavity. The system is dressed such that there is collectively no interaction until a target total photon number is reached across multiple distinct modes, at which point they interact strongly. The microwave cavity features 9 modes with photon lifetimes of $\sim 2$ ms coupled to a superconducting transmon circuit, forming a multimode circuit QED system with single photon cooperativities of $\sim10^9$. We generate multimode interactions by using cavity photon number resolved drives on the transmon circuit to blockade any multiphoton state with a chosen total photon number distributed across the target modes. We harness the interaction for state preparation, preparing Fock states of increasing photon number via quantum optimal control pulses acting only on the cavity modes. We demonstrate multimode interactions by generating entanglement purely with uniform cavity drives and multimode photon blockade, and characterize the resulting two- and three-mode W states using a new protocol for multimode Wigner tomography. 
\end{abstract}
\maketitle

Engineering an interacting many body system requires control of many particles distributed across space or over multiple sites. Recently, superconducting circuit quantum electrodynamics (cQED) has emerged as a platform for creating photonic fluids and materials~\cite{carusotto2020photonic}. 
Circuits can be used to impart effective mass and interactions to microwave photons, creating exotic systems in which the band structure and interactions can be precisely engineered and probed at the single site and single photon (particle) level~\cite{ma2019dissipatively,kollar2019hyperbolic}.
In addition to planar circuits, 3D superconducting microwave cavities have also emerged as a powerful resource for quantum information science and quantum optics, possessing the longest coherence times demonstrated in cQED~\cite{devoret2013outlook, romanenko2020three}. These cavities are interfaced with superconducting circuits to perform gate operations between cavity modes~\cite{rosenblum2018cnot, chou2018deterministic, gao2019entanglement}, and quantum error correction using a variety of bosonic codes~\cite{ofek2016extending,hu2019quantum,campagne2019stabilized}. While the bulk of previous work has been performed in single or two-mode 3D cQED systems, there have also been efforts to create multimode bosonic systems for quantum simulation~\cite{owens2018quarter}, quantum information processing~\cite{naik2017random}, and multimode quantum optics~\cite{sundaresan2015beyond} in cQED, as well as for cavity electromechanics~\cite{pechal2018superconducting,sletten2019resolving}. 
Here, we realize a novel bosonic many body system in a 3D multimode cavity with photons acting as particles, spectrally distinct high quality factor modes (65-95 million) acting as sites, and interactions mediated by a superconducting transmon circuit via multimode cQED. 

A pure N-body interaction is one in which the system remains linear until the $\mathrm{N^{th}}$ quantum is added to the system. This interaction has been engineered on a single cavity mode by using a dispersively coupled transmon qubit for photon blockade~\cite{bretheau2015quantum}, resulting in quantum dynamics that excluded a given photon number N. The notion of blockade~\cite{bretheau2015quantum,vrajitoarea2020quantum} is intimately connected with Quantum Zeno Dynamics (QZD), where repeated measurements suppress the system's unitary evolution and modify it's coherent dynamics~\cite{facchi2000quantum,facchi2008quantum}. Optical measurements have been used to observe QZD~\cite{signoles2014confined, schafer2014experimental} and generate entanglement~\cite{barontini2015deterministic} in atomic systems. QZD has also been proposed as a tool for quantum state preparation of light using atoms~\cite{raimond2010phase,raimond2012quantum}. 

In this work, we broaden the applications of photon blockade for quantum control by using it to demonstrate universal operations on qudits realized in arbitrary bosonic modes of a multimode cavity. We use quantum optimal control pulses that act exclusively on the cavity mode to prepare all basis Fock states of the qudit, despite their equal frequency spacing~\cite{schirmer2001complete}. Furthermore, we extend the notion of photon blockade to a multimode system and realize a pure N-body interaction across an arbitrary number of sites (modes) by selectively driving the transmon near the transition corresponding to a set of chosen N-photon states. This dressing results in a frequency shift that blockades those states, effectively partitioning the multimode cavity's Hilbert space and resulting in the creation of entangled W-states using only uniform drives on the cavity modes. 

\mathchardef\mhyphen="2D

\begin{figure}[t]
\includegraphics[width=0.4832\textwidth]{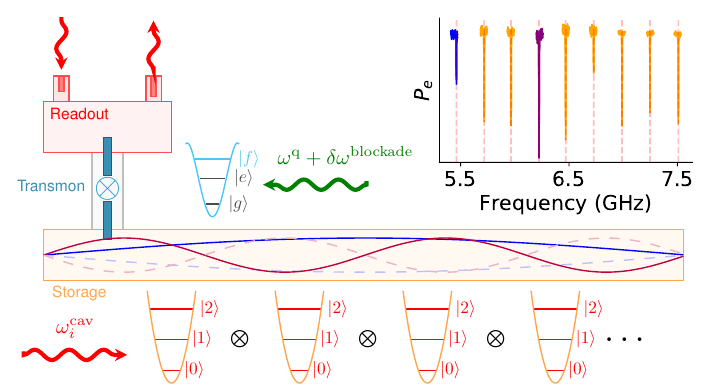}
\caption{\textbf{Schematic of the experimental system}. A multimode rectangular 3D storage cavity and a 3D readout cavity are bridged using a transmon circuit. The multimode Hilbert space designated by the tensor products is controlled by drives that are near the transmon $\ket{g} \mhyphen \ket{e}$ and cavity frequencies,  all input through the readout port. The inset shows the frequency spectrum of the storage cavity with evenly spaced modes from 5.4 to 7.5 GHz, obtained by sweeping the cavity drive near the storage mode frequencies and measuring the population using the transmon.} 

\label{Fig1}
\end{figure}

\begin{figure}[t]
\includegraphics[width=0.4832\textwidth]{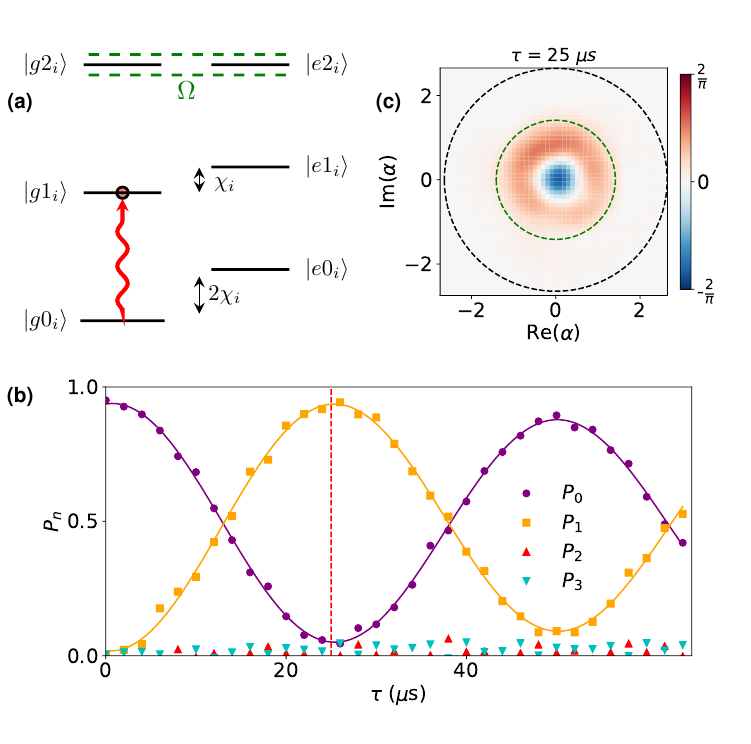}
\caption{\textbf{A qubit implemented in any cavity mode using photon blockade}. (a) Energy level diagram in a rotating frame with an applied blockade drive at Rabi frequency $\Omega$ resonant with $\ket{g2_i} \leftrightarrow \ket{e2_i}$. Transitions at other photon numbers are off resonant by $(n-2)|\chi|$, and only experience Stark shifts. 
(b) Population of each Fock state ($P_n$) over time ($\tau$) for a constant cavity drive ($\epsilon$) applied to cavity mode 3 ($\nu_3 = 6.223$ GHz), in the presence of a photon blockade at $n=2$ with $\epsilon/(2\pi) = 10~\mathrm{kHz} \ll \Omega/(2\pi) = 107~\mathrm{kHz} \ll |\chi_3|/(2\pi)=1.136$ MHz.  
(c) Wigner tomography at $\tau = 25~\mu$s demonstrating the preparation of  $\ket{1}$ with $\mathcal{F} = \mathrm{Tr}[\rho_m\rho_t] = 0.967 \pm 0.024$, after accounting for measurement error by normalizing by the measured fidelity of the vacuum state.}
\end{figure}

\begin{figure*}[t]
\includegraphics[width=0.9\textwidth]{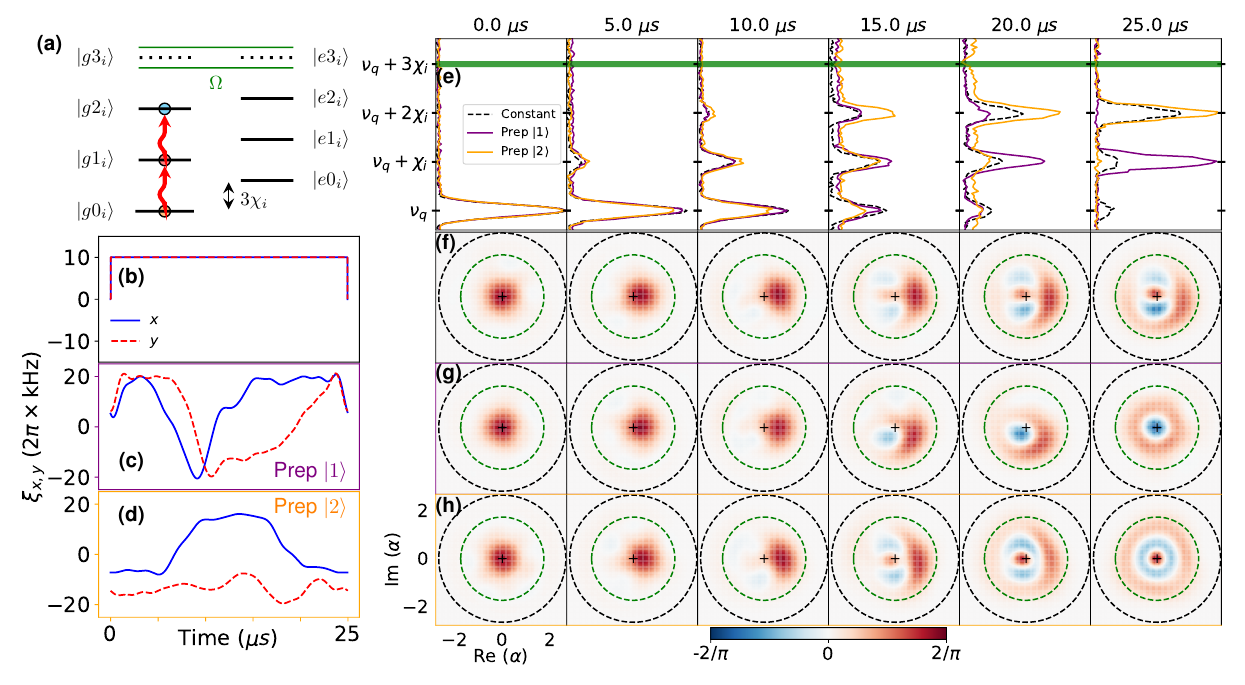}
\caption{\textbf{Universal control of a qudit using optimal control}. (a) Energy level diagram for photon blockade of $n=3$. The blockade drive with Rabi strength $\Omega$ resonant with $\ket{g3_i} \mhyphen \ket{e3_i}$ causes those two levels to hybridize and shift by $\pm \Omega$. A resonant cavity drive leads to transitions in the qutrit formed by $\ket{0},\ket{1},\ket{2}$, with $\ket{3}$ remaining off-resonant and unoccupied.
(b)-(d) The two quadratures of the pulses (I and Q) for a uniform cavity drive and the two optimal control pulses that prepare $\ket{1}$ and $\ket{2}$ from $\ket{0}$ in mode 3. The optimal control pulses are generated using the GrAPE algorithm.
(e) Corresponding spectroscopy of the population evolution over time for the 3 pulses above. Dashed lines indicate the qubit frequencies corresponding to $n=0-3$ photons in mode 3. The thick horizontal green line near the top of the plots marks the frequency of the blockade tone.
(f)-(h) Wigner tomography over time of the uniform drive and optimal control pulses shown in (b)-(d), in the same order. The dashed green circle indicates the location (in phase space) of the blockade drive. The blockade acts as a wall that constrains the region of allowed occupation, with the states prepared by interference from the resulting reflections. }
\end{figure*}
Our system consists of a multimode rectangular waveguide cavity which supports numerous ultralow loss modes on one physical device that are manipulated with multimode cQED via integration with a superconducting transmon circuit, as shown schematically in Fig.~1. We use the $\mathrm{TE_{10n}}$ modes of the cavity to realize a multimode quantum memory. The cavity is fabricated from high-purity (5N5) aluminium using the flute method~\cite{chakramoriani2020}, a design which naturally mitigates seam loss and allows for high quality factors (65-95 million). The mode spectrum is controlled by the cavity length and profile. The second smallest dimension---the cavity width, sets the frequency of the lowest mode ($\sim5.4$ GHz), with higher modes having an increasing number of antinodes along the length. The width is tapered to engineer a frequency spacing of $\sim250$ MHz between the modes (inset of Fig.~1). The modes of the storage cavity are dispersively coupled to a transmon circuit, a nonlinear oscillator whose ground and first excited states implement a qubit with a transition frequency of $\omega_q/(2\pi) = 4.99$ GHz. The transmon is placed near one end of the rectangular waveguide, where all the modes have sufficiently strong zero point fields to couple to the antenna formed by one of the transmon capacitor pads, with resonant dipole interaction strengths ranging from $g_m/(2\pi) = 50-125$ MHz. The second capacitor pad is coupled to a smaller rectangular waveguide cavity, whose fundamental mode ($7.79$ GHz) is strongly coupled to the microwave line ($Q_c = 15\mathrm{k}$). This cavity is used to readout the state of the transmon and, in turn, the storage cavity modes. In the dispersive limit, the Hamiltonian of the multimode cavity QED system realized by the storage cavity and transmon is:
\begin{equation}
\begin{split}
H &=   \omega_q \ket{e}\bra{e} + \sum_{m=0}^{N-1}\{\omega_m a_m^{\dagger} a_m + \chi_m a_m^{\dagger} a_m \ket{e}\bra{e} \, \\ 
&+ \frac{k_m}{2} a_m^{\dagger} a_m (a_m^{\dagger} a_m - 1)\} + \sum_{n \neq m} k_{mn} a_m^{\dagger} a_m a_n^{\dagger}a_n ,
\end{split}
\label{eqn1}
\end{equation}
where $\omega_m$ are the storage cavity mode frequencies and $\chi_m$ are the dispersive shifts -- the qubit Stark shifts per added photon in storage cavity mode $m$ (indexed starting from 0). The coupling to the transmon causes the cavity modes to inherit Kerr nonlinearities, both a self-Kerr shift within a single mode ($k_m$), and a cross-Kerr interaction between modes ($k_{mn}$). The increased detuning from the transmon causes the dispersive shift to decrease with mode number, despite the increase in the interaction strength, with $\chi_m/(2\pi)$ ranging from $-2.39$ to $-0.55$ MHz for the first 9 storage modes ($N = 9$) that are used for operations (see supplementary information). The storage cavity mode lifetimes ($T_1^m\sim 2$ ms), dephasing times ($T_2^m \sim 2-3$ ms)~\cite{chakramoriani2020}, and qubit coherence times ($T^q_1, T^q_2  = 86 \pm 6, 58 \pm 4~\mu$s) result in a system with very high cooperativity ($\sim 10^9$), in which qubit transitions corresponding to different photon numbers in any cavity mode are well resolved~\cite{schuster2007resolving}.
The dispersive interaction can be used for universal control of the cavity states by combining cavity displacements with phase gates performed using photon number selective drives (SNAP) on the transmon, as has been demonstrated previously for a single cavity mode~\cite{heeres2015cavity,heeres2017implementing}

\begin{figure*}[t]
\includegraphics[width=0.975\textwidth]{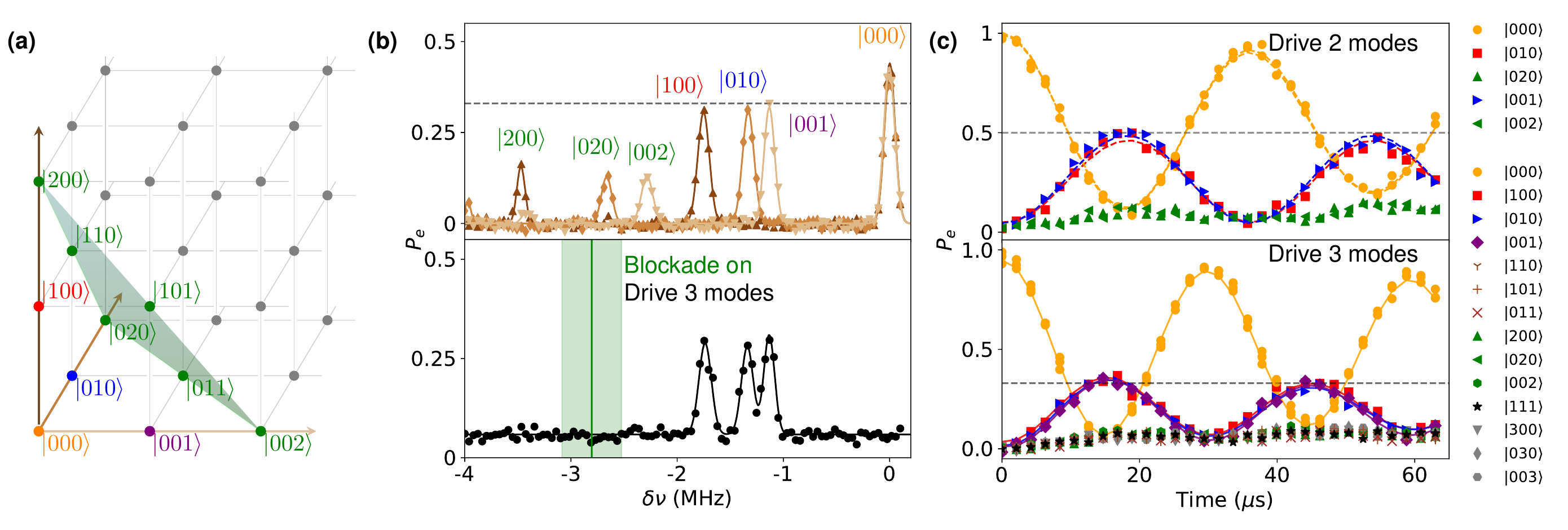}
\caption{\textbf{Generating multimode N-body interactions via photon blockade}. (a) Diagram showing the 3D space of three-mode Fock states with at most 2 photons in each mode. Applying a sufficiently strong single blockade tone at the mean qubit frequency corresponding to the $\ket{011}$, $\ket{101}$, and $\ket{110}$ levels effectively blockades the entire plane of states shown in green. Starting from vacuum ($\ket{000}$), the modes are restricted to the subspace below the plane when the cavity drives are sufficiently weak. This multimode blockade generates a pure M-site N-body interaction. Colors are chosen to correspond to the same states across diagrams.
(b) Qubit spectroscopy measurements. Top panel: coherent states prepared in three modes $(2,3,4)$ with different dispersive shifts, with the same cavity drive strength and duration used below, resulting in occupation of states with more than 1 photon. Bottom panel: a three-mode W-state prepared by driving the cavity modes simultaneously in the presence of the multimode blockade interaction for the appropriate time. 
(c) Population over time for two- (top) and three-mode (bottom) blockades with constant, equal cavity drive strengths. The results are oscillations between multimode entangled states and the ground state. The increase in population of the blockaded states is due to leakage and subsequent decay arising from participation of the transmon. }
\end{figure*}

We use qubit drives to induce N-body interactions that exclude a subset of the multimode Hilbert space, enacting multimode control through a combination of blockade drives near the qubit frequency, and resonant cavity drives. A schematic of the energy levels of the coupled qubit-cavity system for a single cavity mode ($i$) is shown in Fig.~2(a). A coherent qubit drive with Rabi drive strength $\Omega \ll \chi_i$ at frequency $\omega_q + 2\chi_i$ selectively couples the states $\ket{g2_i}$ and $\ket{e2_i}$, causing them to hybridize, and shifting their frequencies by $\pm \Omega$. This takes the $\ket{1} \mhyphen \ket{2}$ transition off resonance from the $\ket{0} \mhyphen \ket{1}$ transition, so that a mode starting from the ground state will undergo a Rabi oscillation between $\ket{0}$ and $\ket{1}$ in the presence of a cavity drive with strength $\epsilon \ll \Omega$, as shown in Fig.~2(b). In phase space, this can be interpreted as the cavity state ``bouncing" off the circle of radius $\sqrt{2}$ defined by the blockade (green circle in Fig.~2(c)). At the time that corresponds to a $\pi$-pulse of the Rabi oscillation, the interference pattern produced by bouncing off of the blockade circle prepares $\ket{1}$, as demonstrated by the cavity Wigner function in Fig.~2(c). Although the transmon does not appear to participate directly in the state preparation, its decay and dephasing still limit the state's fidelity. This is due to a combination of leakage to the dressed $\ket{2}$ state with subsequent decay via the transmon ($\epsilon/(\Omega^2 T^q)$), and Purcell decay of the $\ket{1}$ state from its dressing with the transmon by the off-resonant blockade tone ($\Omega^2/(\epsilon\chi^2 T^q)$), where $T_q$ is the qubit decoherence time that is the minimum of $T^1_q$ and $T^2_q$. Optimizing the cavity drive strength results in a minimum infidelity of $\sim1/(\chi T^q)$, which is comparable to that of SNAP gates in the absence of cavity decay.

The Rabi oscillations produced by driving the cavity while blockading $\ket{2}$ allow us to create arbitrary superpositions of $\ket{0}$ and $\ket{1}$. This universal control can be extended to higher dimensional qudits, despite the nearly identical transition frequency of the $d$ levels, as shown theoretically in~\cite{schirmer2001complete}. Similar to the case of $d=2$ above, we implement a qudit in mode $i$ by driving the $\ket{gd_i}\leftrightarrow\ket{ed_i}$ transmon transition to blockade $\ket{d_i}$, as illustrated in the energy level diagram in Fig.~3(a) for a qutrit obtained by blockade of $\ket{3_i}$. We demonstrate universal qutrit control through shaped pulses obtained using the gradient ascent pulse engineering (GrAPE) algorithm~\cite{khaneja2005optimal}, implemented using the optimal control package developed in~\cite{leung2017speedup}. The drift Hamiltonian used to obtain the pulses is in a doubly rotating frame at the cavity and blockade drive frequency, and includes the dispersive shift and self-Kerr interaction of the mode of interest. The control pulses are constrained to have a length of 25 $\mu$s, with maximum amplitudes $\epsilon \ll \Omega$, and pulse bandwidth limited to $\pm \chi/2$ around the cavity frequency. The real and imaginary quadratures of the pulse envelopes used to prepare $\ket{1}$ and $\ket{2}$ from the vacuum state with a blockade at $\ket{3}$ are shown in Fig.~3(c) and (d), respectively. 
A uniform cavity drive (Fig.~3(b)) produces a superposition of $\ket{0}, \ket{1}$, and $\ket{2}$, shown by the dashed black qubit spectroscopy curve in Fig.~3(e) and the Wigner functions in Fig.~3(f). The optimal control pulses shown in Fig.~3(c) and (d) prepare $\ket{1}$ and $\ket{2}$ with fidelities of $0.953 \pm 0.022$ and $0.965 \pm 0.022$, respectively.
The corresponding Wigner function evolutions are shown in Fig.~3(g) and (h). 
The measurements presented here were performed on cavity mode 3, with $\nu_3 =  6.223$ GHz and $\chi_3 = -1.136$ MHz. These methods allow for universal control of qutrits in any of the cavity modes, and can be extended to higher dimensional qudits.

\begin{figure}[t]
\includegraphics[width=0.5\textwidth]{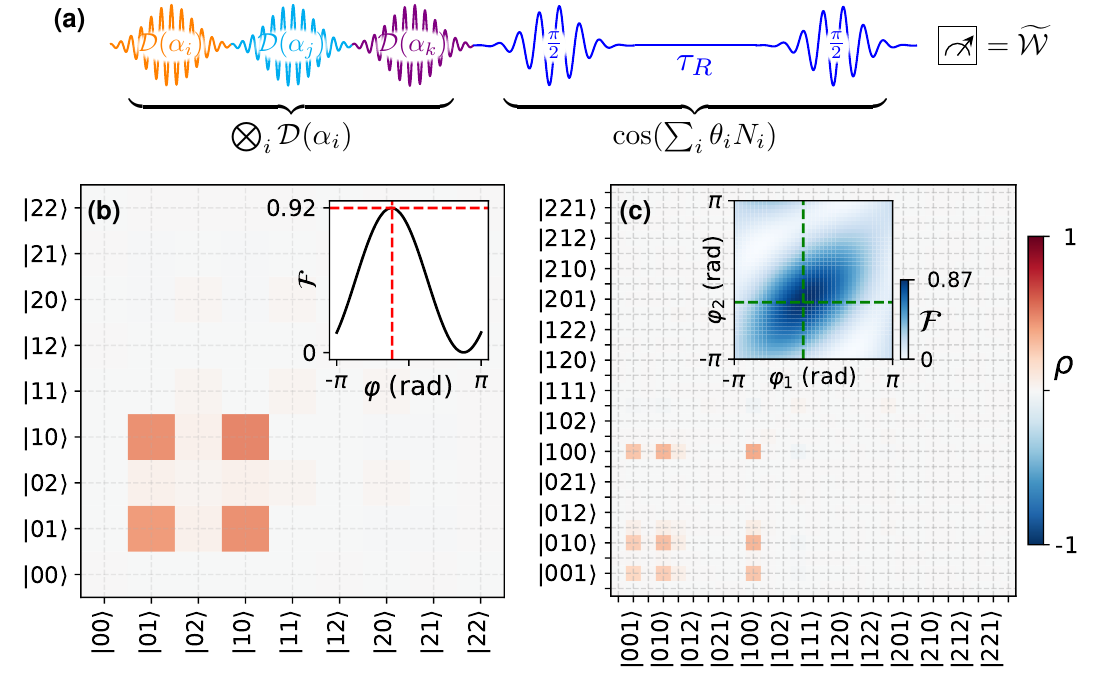}
\caption{\textbf{Multimode Wigner tomography}. (a) The pulse sequence involves displacements of each of the modes followed by a qubit Ramsey measurement with a wait time $\tau$ -- acting as an approximate joint parity measurement, and resulting in the overall measurement of a generalized multimode Wigner operator. This operator is measured for a set of known displacements of each of the modes, from which the density matrix ($\rho$) is obtained. 
(b) Real part of the reconstructed $\rho$, for a two-mode entangled state of modes 3 and 4, and (c) for a three-mode entangled state of modes 2, 3 and 4. The phases of the superpositions are extracted by maximizing the inner product of the measured $\rho$ and an ideal W-state with free phases (insets). The gauge freedom in the phase definition for each mode is used to make the density matrices real.}
\end{figure}

We extend photon blockade to implement N-body interactions across multiple modes and utilize them for quantum state preparation. Creating these interactions requires shifting the energies of multiple multimode states. We accomplish this by choosing a single drive tone that blockades a set of multimode states whose dispersive shifts are sufficiently close in frequency. This creates a hyper-plane of blockaded states with the same total photon number, whose energies are all shifted, defining the multimode N-body interaction. This hyper-plane effectively partitions the space of states that are accessible by cavity drives weaker than the interaction strength. This could also be accomplished by using multiple blockade drives at different frequencies, but we will focus on the single-tone case.

This plane is shown in Fig.~4(a) for 2 photons and 3 modes, whose boundary contains the blockaded states for the two-mode case. The interaction is implemented with a blockade drive offset from the qubit frequency by the average dispersive shifts of combinations of adding 2 photons, i.e., $\delta\nu_b =\chi_i + \chi_j$ for 2 modes ($i,j$), and $\delta\nu_b =(2\chi_i + 2\chi_j+2\chi_k)/3$ for 3 modes ($i,j,k$). As in the single mode case, the blockade drive must be weak enough to satisfy $\Omega \ll \delta\nu_b - |\chi_{i,j,k}|$ so as to not target states with a different total photon number. On the other hand, the drive must also be strong enough to simultaneously blockade all the multimode states with a total of 2 photons. For example, in the two-mode case, blockading $\ket{20}$, $\ket{02}$, and $\ket{11}$ requires $\Omega > |\delta\chi| = |\chi_i-\chi_j|$. In our system, these criteria can be satisfied for all mode pairs that are nearest neighbours in frequency (see supplementary information). Subsequently, driving all cavity modes with sufficiently weak drives ($\epsilon < \Omega$) in the presence of the blockade interaction results in constrained multimode dynamics within the space of 0 and 1 photon when starting from the vacuum state. In particular, driving the modes with equal strengths for an appropriate time results in the preparation of W-states, $(\ket{10} + e^{i\phi}\ket{01})\sqrt{2}$ and $(\ket{100} + e^{i\phi_1}\ket{010} + e^{i\phi_2}\ket{001}) / \sqrt{3}$ for the two- and three-mode cases, respectively.

We first demonstrate this interaction with qubit spectroscopy. In the absence of the blockade drive, independent weak drives in each mode produce coherent states with occupation in the two photon levels $\ket{200},\ket{020}$ or $\ket{002}$, as shown in Fig.~4~(b, top). However, in the presence of the blockade, the states with more than 1 photon have no occupation, as shown in Fig.~4(b, bottom). Here, 3 modes are driven simultaneously with equal strengths for the time required to prepare a W-state, resulting in peaks only at $\ket{100}, \ket{010}$ and $\ket{001}$. Using the peak heights from spectroscopy, we measure the population oscillations as a function of time in Fig.~4(c) when 2 (top) or 3 (bottom) modes are simultaneously driven. We see that most of the population remains in the 0 and 1 photon multimode subspace, and that the populations are consistent with a W-state at the time when the vacuum state is emptied.

To completely characterize the entangled states prepared in this system, we perform multimode Wigner tomography using the protocol shown in Fig.~5(a). 
A photon in mode $m$ causes the qubit to accrue a known phase $\theta_m = 2\pi\chi_m\tau_R$, where $\tau_R$ is the wait time during a qubit Ramsey measurement.
For modes with different $\chi$'s, the phase accrues at different rates. Prior schemes have addressed this by using higher transmon levels to perform a joint parity measurement~\cite{wang2016schrodinger}, $\hat{\Pi} = e^{i\pi\sum_{m}{N_m}}$, where $N_m$ is the number of photons in mode $m$. Here, we instead perform tomography using a generalization of a joint parity measurement, which forgoes the need for dispersive shift engineering or the use of higher transmon levels.
A qubit Ramsey sequence with idle time corresponding to phase angles $\theta_m$ measures the operator $\widetilde{\Pi} = \cos(\sum_{m}{\theta_m N_m})$. Adding cavity displacements prior to the Ramsey measurement ($U = \bigotimes_{m}\mathcal{D}(\alpha_m)$) measures the generalized Wigner function $\widetilde{\mathcal{W}} = U\widetilde{\Pi} U^{\dagger}$, which reduces to the usual joint Wigner function for $\theta_m = \pi$. This function is measured for a set of mode displacements, whose results are inverted to reconstruct the multimode density matrix (see supplementary information). The real parts of the density matrices obtained from a reconstruction for W-states of two and three modes are shown in Fig.~5(b) and (c), with state fidelities of $\mathcal{F} = \mathrm{Tr}[\rho_M\rho_T] =  0.918 \pm 0.012$ and $0.864 \pm 0.014$, respectively.

In summary, we have used photon blockade to realize a multimode N-body interaction in arbitrary modes of a multimode cQED system. We demonstrate the ability to prepare Fock states of increasing photon number in any mode, using a new scheme that involves optimal control pulses acting only on the cavity in the presence of a blockade drive. We also prepare entangled W-states of photons in up to three modes with a scheme that easily generalizes to larger numbers of modes. To characterize the entanglement of the prepared states, we develop a new protocol for multimode Wigner tomography that does not require engineering the dispersive interaction. In future work, more exotic interactions and partitions of the multimode Hilbert space can be implemented by applying multiple blockade tones. Additionally, states prepared and evolved in separate blockade subspaces can be made to interfere with each other, thereby realizing a multimode analog of phase space tweezers~\cite{raimond2010phase}.
The multimode cQED system demonstrated in this work advances the state-of-the-art in terms of number of modes, lifetimes, and cooperativities, and is a promising new platform for exploring interactions and many-body states of microwave photons relevant for quantum optics, quantum simulation, and quantum error correction.

\begin{acknowledgments} 
We thank Thomas Propson, Yao Lu, Ankur Agrawal, Tanay Roy, and Jon Simon for useful discussions. We acknowledge support from the Samsung Advanced Institute of Technology Global Research Partnership. This work was also supported by ARO Grants W911NF-15-1-0397 and W911NF-18-1-0212, ARO MURI grant W911NF-16-1-0349, AFOSR MURI grant  FA9550-19-1-0399, and the Packard Foundation (2013-39273). This work is funded in part by EPiQC, an NSF Expedition in Computing, under grant CCF-1730449. We acknowledge the support provided by the Heising-Simons Foundation. D.I.S. acknowledges support from
the David and Lucile Packard Foundation. This work was partially supported by the University of Chicago Materials Research Science and Engineering Center, which is funded by the National Science Foundation under award number DMR-1420709. Devices were fabricated in the Pritzker Nanofabrication Facility at the University of Chicago, which receives support from Soft and Hybrid Nanotechnology Experimental (SHyNE) Resource (NSF ECCS-1542205), a node of the National Science Foundation’s National Nanotechnology Coordinated Infrastructure.
\end{acknowledgments}

\bibliography{thebibliography}
\end{document}


\mathchardef\mhyphen="2D
\title{Multimode photon blockade: Supplementary Information} 
\date{\today}
\maketitle

\section{Cryogenic setup and control instrumentation}
\begin{figure}[h]
  \begin{center}
    \includegraphics[width= 0.5\textwidth]{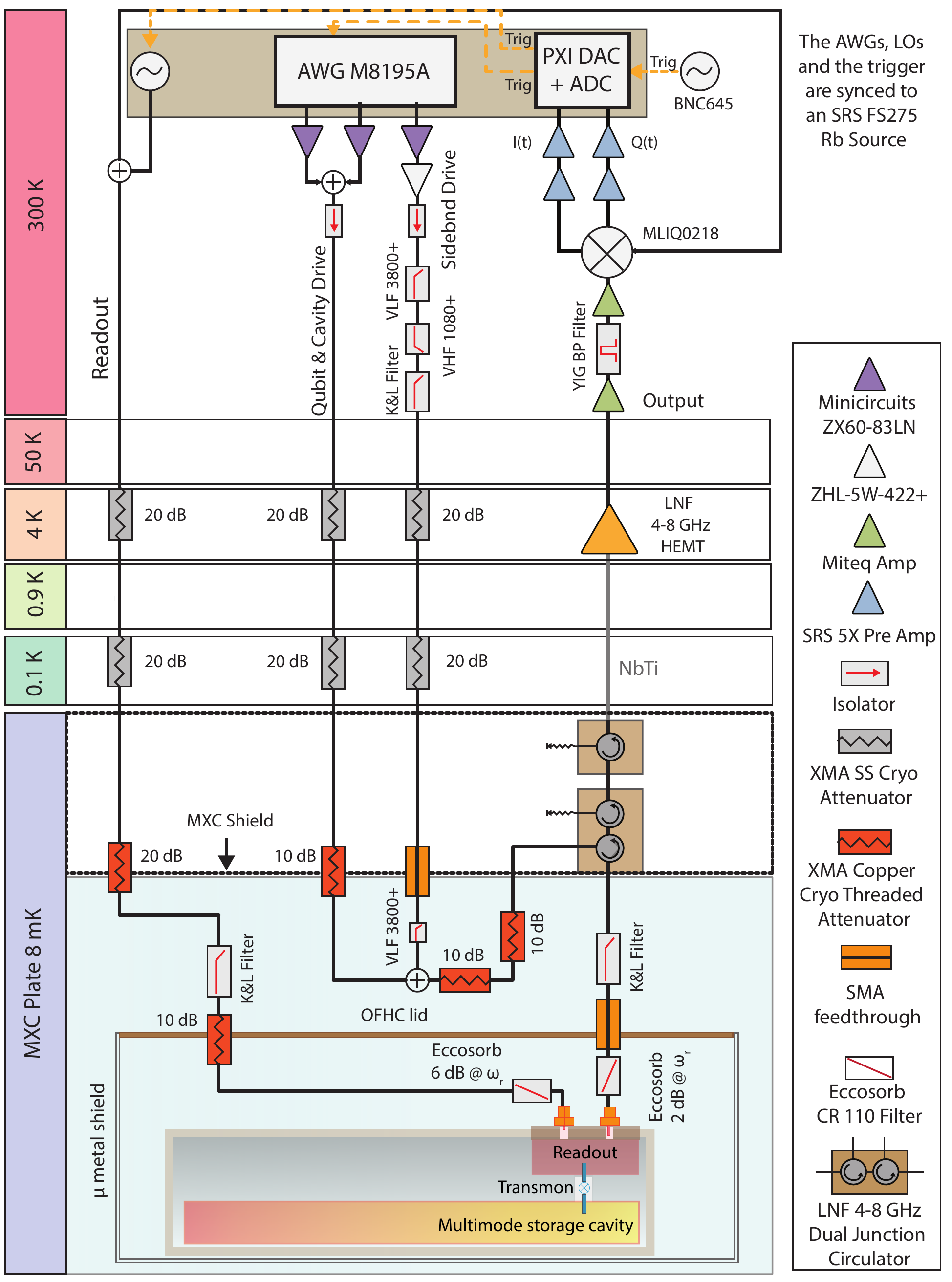}
    \caption{Schematic of the cryogenic setup, microwave wiring and filtering, and control instrumentation.}
  	\label{wiring diagram}
  \end{center}
\end{figure} 
The multimode cavity device is heat sunk to an OFHC copper plate connected to the base stage of a Bluefors LD-400 dilution refrigerator (7-8 mK). The sample is surrounded by a can containing two layers of $\mu$-metal shielding, with the inside of the inner layer connected to a can made out of copper shim that is attached to the copper can lid and painted on the inside with Berkeley black. A schematic of the cryogenic setup, control instrumentation, and device wiring is shown in SFig.~\ref{wiring diagram}. The device is machined from a single piece of 5N5 aluminium and consists of a readout cavity and a multimode storage cavity fabricated using the flute method described in~\cite{chakramoriani2020}. The cavities are bridged by a 3D transmon circuit whose fabrication is detailed in the next section. All controls are performed through the readout cavity, by driving at the qubit and storage mode frequencies. The pulses are directly digitally synthesized using a 4-channel, 64 GSa/s arbitrary waveform generator (Keysight M8195A). The combined signals are sent to the device after being attenuated at each of the thermal stages, as shown in SFig.~\ref{wiring diagram}. The transmitted signal from the readout resonator passes through three cryogenic circulators (thermalized at the base stage) and is amplified using a HEMT amplifier (anchored at 4 K). Outside the fridge, the signal is filtered (tunable narrow band YIG filter with a bandwidth of 80 MHz) and further amplified. The amplitude and phase of the resonator transmission signal are obtained through a homodyne measurement, with the transmitted signal demodulated using an IQ mixer and a local oscillator at the readout resonator frequency. The homodyne signal is amplified (SRS preamplifier) and recorded using a fast ADC card (Keysight M3102A PXIe 500 MSa/s digitizer).

\section{Fabrication of the transmon ciruit}
The transmon qubit was fabricated on a $\SI{430}{\micro \meter}$ thick C-plane (0001) sapphire wafer with a diameter of $\SI{50.8} {\milli \meter}$. The wafer was cleaned with organic solvents (Toluene, Acetone, Methanol, Isopropanol, and DI water) in an ultrasonic bath to remove contamination, then annealed at $\SI{1200} {\degreeCelsius}$ for 1.5 hours. Prior to film deposition, the wafer underwent a second clean with organic solvents (Toluene, Acetone, Methanol, Isopropanol, and DI water) in an ultrasonic bath. The junction was made out of aluminum using a combination of optical and electron-beam lithography. The base layer of the device, which includes the capacitor pads for the transmon, consists of $120$ nm of Al deposited via electron-beam evaporation at $1 \si{\angstrom} /s$. The features were defined via optical lithography using AZ MiR 703 photoresist and exposure by a Heidelberg MLA150 Direct Writer. The resist was developed for 1 minute in AZ MIF 300 1:1. The features were etched in a Plasma-Therm inductively coupled plasma (ICP) etcher using chlorine based etch chemistry (30 sccm $\text{Cl}_2$, 30 sccm $\text{BCl}_2$, 10 sccm Ar). This was followed by a second layer of optical patterning and thermal evaporation of 50 nm of Au for the alignment marks used for ebeam lithography. The resist was subsequently removed by leaving the wafer in $80^{\circ}$C N-Methyl-2-pyrrolidone (NMP) for 4 hours.
The junction mask was defined through electron-beam lithography of a bi-layer resist (MMA-PMMA) in the Manhattan pattern using a Raith EBPG5000 Plus E-Beam Writer, with overlap pads that allow for direct galvanic contact to the optically defined capacitors. The resist stack was developed for 1.5 minutes in a solution of 3 parts IPA and 1 part DI water. Before deposition, the overlap regions on the pre-deposited capacitors were milled \textit{in-situ} with an argon ion mill to remove the native oxide. The junction was then deposited with a three step electron-beam evaporation and oxidation process. First, an initial $35$ nm layer of Al was deposited at $1$ nm/s at an angle of $29^\circ$ relative to the normal of the substrate, azimuthally parallel to one of the fingers in the Manhattan pattern. Next, the junction was exposed to $20$ mBar of a high-purity mixture of $\mathrm{Ar}$ and $\mathrm{{O}_2}$ (80:20 ratio) for 12 minutes to allow the first layer to grow a native oxide. Finally, a second $120$ nm layer of Al was deposited at $1$ nm/s at the same $29^\circ$ angle relative to the normal of the substrate, but azimuthally orthogonal to the first layer of Al. After evaporation, the remaining resist was removed via liftoff in $80^{\circ}$C NMP for 3 hours, leaving only the junction directly connected to the base layer. After both the evaporation and liftoff, the device was exposed to an ion-producing fan for 30 minutes to avoid electrostatic discharge of the junction.

\section{Calibration of the multimode Hamiltonian}
\begin{figure}[t]
  \begin{center}
    \includegraphics[width= 0.5\textwidth]{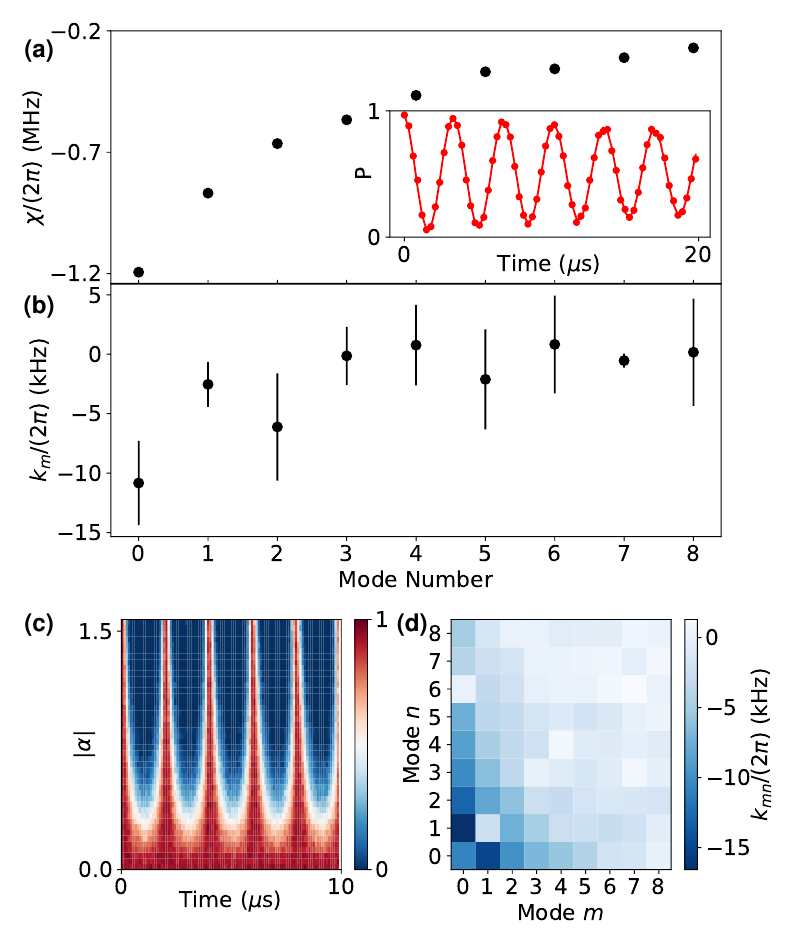}
    \caption{\textbf{Calibrations of multimode cavity dispersive shift and self-Kerr interactions}.
    (a) Dispersive shift calibration for each of the manipulable modes. The measurement is performed by placing a photon in a mode, followed by a qubit Ramsey and fitting to the resulting oscillation frequency. An example is shown in the inset.
    (b) Self-Kerr calibration for each of the modes. The measurement is performed through cavity Ramsey and fitting to the resulting spectrum vs. time and the magnitude $|\alpha|$ of the cavity displacement. 
    (c) Self-kerr data for cavity mode number 3. 
    (d) Cross-shift between mode pairs. The measurement is performed by placing a photon in mode $m$, then sweeping the cavity frequency when probing the 0 to 1 photon peak of mode $n$.
    }
  	\label{cavity_parameter_calibrations}
  \end{center}
\end{figure}

The Hamiltonian of the multimode cavity QED system realized by the transmon and the storage modes is:
\begin{equation}
\begin{split}
H &=   \omega_q \ket{e}\bra{e} + \sum_{m=0}^{N-1}\{\omega_m a_m^{\dagger} a_m + \chi_m a_m^{\dagger} a_m \ket{e}\bra{e} \, \\ 
&+ \frac{k_m}{2} a_m^{\dagger} a_m (a_m^{\dagger} a_m - 1)\} + \sum_{n \neq m} k_{mn} a_m^{\dagger} a_m a_n^{\dagger}a_n ,
\end{split}
\label{seqn1}
\end{equation}
where $\omega_q$ is the frequency of the transmon $\ket{g} \mhyphen \ket{e}$ transition, $\omega_{m}$ the memory mode frequencies, $\chi_m$ the dispersive shifts, $k_m$ the self-Kerr shift of each mode, and $k_{mn}$ the cross-Kerr interactions between the modes. The value of $\omega_q$ is obtained through a standard Ramsey measurement on the transmon. The $\chi_m$ are initially calibrated by performing qubit spectroscopy with a resolved pulse swept near the qubit frequency, following a coherent drive at the cavity frequency. The $\chi_m$ are then determined more precisely with a Ramsey experiment on the transmon $\ket{g} \mhyphen \ket{e}$ transition after the addition of a photon in the cavity mode, as shown in SFig.~\ref{cavity_parameter_calibrations}(a). The photon is added to the cavity either by initializing the transmon in $\ket{f}$ and then driving the $\ket{f0}-\ket{g1_m}$ transition, or by performing a Rabi oscillation on the cavity in the presence of a blockade at $n=2$, as described in the main text. The $k_m$ of the cavity modes are obtained by performing a cavity Ramsey experiment, with the measured values shown in SFig~\ref{cavity_parameter_calibrations}(b). This experiment is conducted by varying the time ($\tau$) between two coherent cavity pulses (with the the phase of the second cavity pulse advanced by $2\pi\nu_R\tau$) and subsequently measuring the population in $\ket{0}$ using a resolved transmon $\pi$ pulse. The magnitude of the coherent state $\alpha$ injected in the cavity is also swept, and the resulting data is fit to the expression:
$P_0(t) = |\exp(-\alpha^2) \sum_{n}^{} \frac{1}{n!} \alpha^{2n} \exp(-itn(\omega_m + k_m n/2))|^2$, as shown in SFig.~\ref{cavity_parameter_calibrations}(c) for cavity mode 3.
The cross-Kerrs $k_{mn}$ are obtained by adding a photon to mode $m$ and performing cavity spectroscopy on a different mode $n$. This procedure is also repeated, on the same mode to verify the consistency of the self-Kerr shifts. The values of $k_{mn}$ are shown in SFig.~\ref{cavity_parameter_calibrations}(d). A summary of all measured quantities relevant to the Hamiltonian, as well as Liouvillian terms corresponding to transmon and cavity decoherence and decay, is provided in STable~\ref{parameters_table}.

\begin{table*}
\def\arraystretch{1.3}  
\begin{tabular}{|c|c|c|c|}
\hline
\rowcolor{Gray}
Parameter & Hamiltonian/Liouvillian Term & Quantity & Value \\
\hline
\rowcolor{tmoncolor}
Transmon frequency & $\omega_q \ket{e}\bra{e}$ & $\omega_q/(2\pi)$ & 4.99 GHz \\
\hline
\rowcolor{storagecolor}
Storage cavity frequencies & $\omega_m a_m^{\dagger} a_m$ & $\omega_m/(2\pi)$ & 5.46 - 7.51 GHz \\
\hline
\rowcolor{readoutcolor}
Readout frequency &  $\omega_r a_r^{\dagger} a_r$ & $\omega_r/(2\pi)$ & 7.79 GHz \\
\hline
\rowcolor{readoutcolor}
Readout dispersive shift & $\chi_r a_r^{\dagger} a_r \ket{e}\bra{e}$ & $\chi_r/(2\pi)$ & 1 MHz \\
\hline
\rowcolor{storagecolor}
Storage mode dispersive shifts & $\chi_m a_m^{\dagger} a_m \ket{e}\bra{e}$ & $\chi_m/(2\pi)$ & see SFig.~\ref{cavity_parameter_calibrations} \\
\hline
\rowcolor{storagecolor}
Storage mode self-Kerrs & $\frac{k_m}{2} a_m^{\dagger} a_m (a_m^{\dagger} a_m - 1)$ & $k_m/(2\pi)$ & '' \\
\hline
\rowcolor{storagecolor}
Storage mode cross-Kerrs & $k_{mn}a_m^{\dagger} a_m a_n^{\dagger} a_n $ & $k_{mn}/(2\pi)$ & '' \\
\hline
\rowcolor{tmoncolor}
Transmon $\ket{e} \rightarrow \ket{g}$ relaxation & $\frac{1}{T^q_1}(1+\bar{n}) \mathcal{D}\big[\ket{g}\bra{e}\big]$ & $T^q_1$ & $86 \pm 6~\mu$s  \\
\hline
\rowcolor{tmoncolor}
Transmon $\ket{g}-\ket{e}$ dephasing & $(\frac{1}{T^q_2}-\frac{1}{2T^{q}_{1}}) \mathcal{D}\big[\ket{e}\bra{e}\big]$ & $T^q_2$ & $ 58 \pm 4~\mu$s \\
\hline
\rowcolor{readoutcolor}
Readout linewidth &  $\kappa_r \mathcal{D}[a_r]$ & $\kappa_r/(2\pi)$ & 0.52 MHz \\
\hline
\rowcolor{storagecolor}
Storage mode relaxation & $\frac{1}{T^m_1} \mathcal{D}[a]$ & $T^m_1$ & $\sim 2$ ms, see~\cite{chakramoriani2020} \\
\hline
\rowcolor{tmoncolor}
Transmon thermal population & $\frac{\bar{n}}{T^q_1} \mathcal{D}\big[\ket{e}\bra{g}\big]$ & $\bar{n}$ & $1.2 \pm 0.5$ \% \\
\hline
\rowcolor{storagecolor}
Storage mode dephasing & " &  $T^m_2$ &$\sim 2-3$ ms, see~\cite{chakramoriani2020}\\
\hline

\end{tabular}
\captionsetup{justification=centering}
\caption{\textbf{Multimode cQED system parameters}}
\label{parameters_table}
\end{table*}

\begin{figure}[t]
  \begin{center}
    \includegraphics[width= 0.5\textwidth]{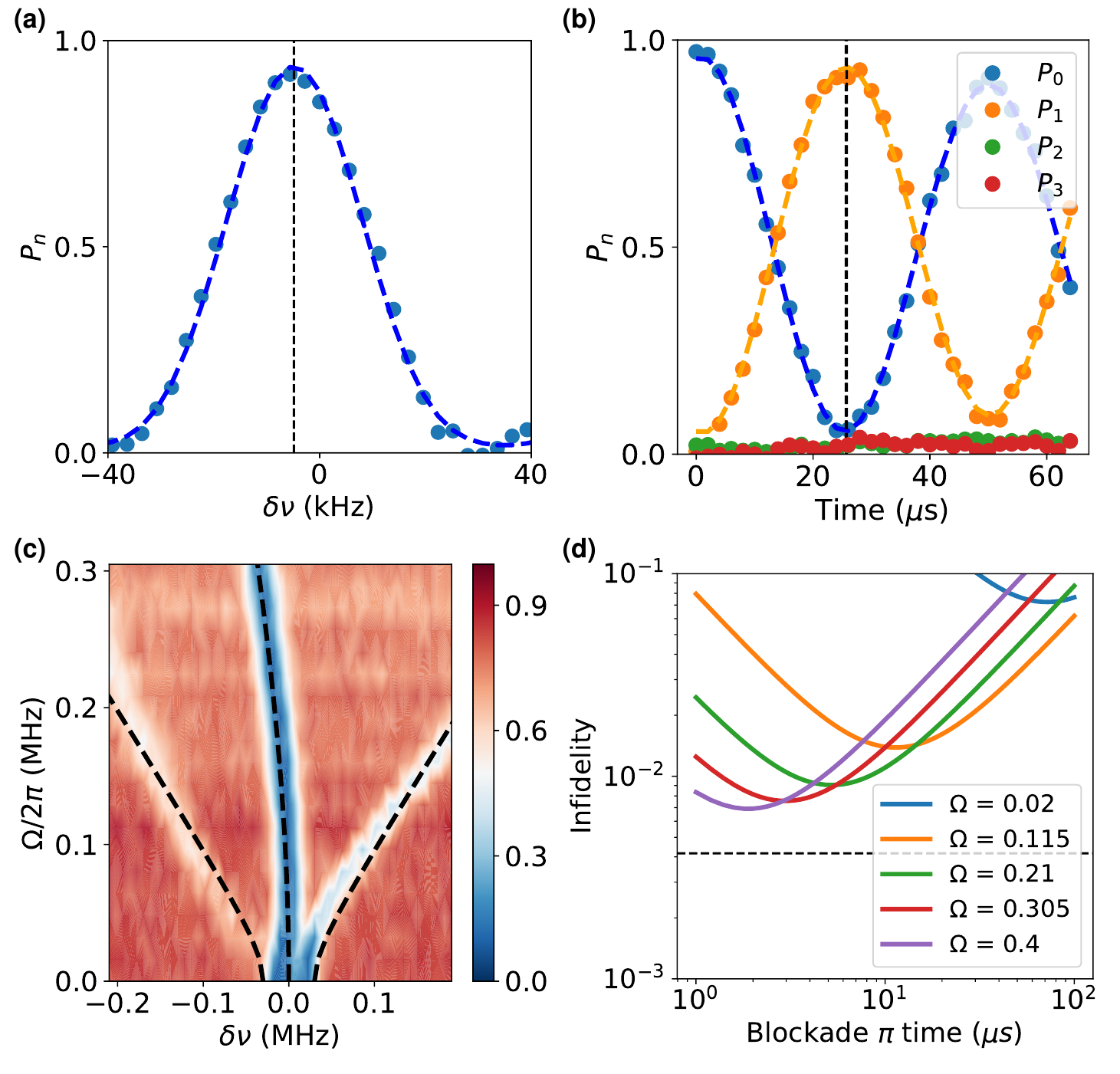}
    \caption{\textbf{Blockade calibrations.}
    (a) Cavity spectroscopy in the presence of the blockade drive. The fitted center frequency is marked by the vertical dashed black line, and indicates the Stark shift from the bare cavity resonance frequency due to the blockade.
    (b) Populations over time produced by a constant cavity drive ($\epsilon/(2\pi) = 10$ kHz) with a blockade drive ($\Omega/(2\pi) = 107$ kHz) at $\ket{2}$, resulting in a Rabi oscillation between the $\ket{0}$ and $\ket{1}$ levels.
    (c) Spectrum showing the energy level splitting as a function of the blockade drive strength $\Omega$. The central vertical blue line shows the $\ket{0} \mhyphen \ket{1}$ transition, which bends due to the Stark shift. The left and right lines show the Rabi split $\ket{1} \mhyphen \ket{2}$ transitions. The corresponding theoretical curves are indicated by the dashed black lines.
    (d) Theoretical blockade infidelity as a function of the the time required for a blockade $\pi$ pulse ($1/(2\epsilon)$), for different blockade drive strengths $\Omega$, as a result of both transmon decoherence and cavity decay.}
  	\label{blockade_calibrations}
  \end{center}
\end{figure}

The minimal description of the dynamics during the blockade of a single mode includes the dispersive coupling between the transmon and the cavity, a Rabi drive on the transmon $\ket{g}$-$\ket{e}$ transition, the self-Kerr of the mode, and the cavity drive. The corresponding Hamiltonian in the frame rotating at the dressed mode and transmon frequencies is:
\begin{equation}
\begin{split}
\hat{H} =~\chi\hat{a}^{\dagger}\hat{a}\ket{e}\bra{e} &+ \frac{\kappa}{2}\hat{a}^{\dagger}\hat{a}\left(\hat{a}^{\dagger}\hat{a}-1\right)  \\
&+ \left\{\Omega(t)\ket{g}\bra{e} + \epsilon(t)\hat{a}+\mathrm{c.c.}\right\}.
\end{split}
\label{blockade_ham}
\end{equation}
To blockade the $|n_0\rangle$ Fock state of a single mode, the transmon is driven at frequency $\omega_q + \chi n_0$. The blockade drive can thus be expressed as $\Omega(t) = \tilde{\Omega}e^{-i\chi n_0 t}$. We make the blockade drive term static through the frame transformation $\hat{U} = e^{-i\chi\ket{e}\bra{e}n_0 t}$, resulting in:
\begin{equation}
\begin{split}
\tilde{\hat{H}} &= \chi\left(\hat{a}^{\dagger}\hat{a}-n_0\right)\ket{e}\bra{e} + \frac{\kappa}{2}\hat{a}^{\dagger}\hat{a}\left(\hat{a}^{\dagger}\hat{a}-1\right)\\
&+ \ \left\{\Omega\ket{g}\bra{e} + \epsilon(t)\hat{a}+\mathrm{c.c.}\right\}.
\label{blockade_ham_rf}
\end{split}
\end{equation}
The blockade is valid in the regime that $\epsilon\sqrt{n_0} < \Omega < \chi$. The first of these conditions prevents leakage to $\ket{\widetilde{g,n_0}},\ket{\widetilde{e,n_0}}$, while the second selectively blockades only the $\ket{g,n_0}\leftrightarrow\ket{e,n_0}$ transition and minimally affects transitions corresponding to other photon numbers. This Hamiltonian can be simplified by individually diagonalizing each photon number subspace ($\ket{g,n}$, $\ket{e, n}$). The blockade drive is resonant with $\ket{g,n_0}\rightarrow\ket{e,n_0}$, splitting those levels by $2\Omega$ and mixing them equally. For levels on either side of $n_0$, the dressing between the ground and excited states is proportional to $\Omega/\left(\chi\left(n-n_0\right)\right)$ to leading order in $\Omega/\chi$. The Hamiltonian can be rewritten in terms of these dressed states as:
\bea 
H &-& \chi\left(\hat{a}^{\dagger}\hat{a}-n_0\right) = \xi(t)\left(\sqrt{n+1}\ket{\widetilde{g,n}}\bra{\widetilde{g,n}+1} + \mathrm{c.c} \right)\nonumber\\
+&\mathlarger{\sum_{n}}&\sqrt{\frac{\chi^2\left(n-n_0\right)^2}{4} + \Omega^2}
\big(\ket{\widetilde{e,n}}\bra{\widetilde{e,n}}-\ket{\widetilde{g,n}}\bra{\widetilde{g,n}}\big)
\eea
In the above, we have dropped the drive terms that couple the dressed ground and excited states, which are off-resonant and suppressed by $\Omega/\chi$. The physics of the blockade can be approximated within a truncated Hilbert space that involves only the dressed transmon ground state, described by the following Hamiltonian:
\begin{widetext}
\begin{equation}
\begin{split}
H &\approx\sum_{n}\left(\left[-\sqrt{\chi^2\left(n-n_0\right)^2/4 + \Omega^2} + \chi(n-n_0)\right]\ket{\widetilde{g,n}}\bra{\widetilde{g,n}} + \xi(t)\left(\sqrt{n+1}\ket{\widetilde{g,n}}\bra{\widetilde{g,n}+1} + \mathrm{c.c}   \right)\right)\\
&\approx \sum_{n}\left(-\frac{\Omega^{2}}{4\chi\left(n-n_0\right)}\ket{\widetilde{g,n}}\bra{\widetilde{g,n}} + \xi(t)\left(\sqrt{n+1}\ket{\widetilde{g,n}}\bra{\widetilde{g,n}+1} + \mathrm{c.c} \right)\right).
\label{reduced_Ham}
\end{split}
\end{equation}
\end{widetext}
We characterize the blockade interaction with the experiments shown in SFig.~\ref{blockade_calibrations}. SFig.~\ref{blockade_calibrations}(a) depicts the Stark shift of the cavity $\ket{0} \mhyphen \ket{1}$ as a result of a blockade at $\ket{2}$. Driving the cavity in the presence of the blockade generates Rabi oscillations that can be used to prepare a photon in the cavity, as shown in (b), with the theoretical infidelity versus the time for a cavity blockade $\pi$ pulse shown in (d). As is evident from the Hamiltonian SEqn.~(\ref{reduced_Ham}), the blockade drive Stark shifts the other cavity Fock states. The resulting blockade spectrum is probed using the following cavity spectroscopy experiment. We first prepare the cavity in $\ket{1}$ through the method described above. We then sweep the frequency of a weak cavity drive tone, and perform spectroscopy as a function of the Rabi amplitude ($\Omega$) of the blockade drive. We observe peaks corresponding to the Stark shifted $\ket{0} \mhyphen \ket{1}$ transition and the two Rabi split $\ket{1} \mhyphen \ket{2}$ transitions, as shown in SFig.~\ref{blockade_calibrations}(c). The Rabi amplitude $\Omega$ is calibrated using transmon Rabi oscillations.

\section{Generation of optimal control pulses in the presence of the blockade drive}

Optimal control pulses were generated with the GrAPE algorithm using the package developed in~\cite{leung2017speedup} using two methods. The first approach used the Hamiltonian in SEqn.~(\ref{blockade_ham_rf}), which is in the frame rotating at the blockade frequency, using a Hilbert space with 2 transmon levels and 5-7 cavity levels,. It includes a fixed transmon blockade drive included in the drift Hamiltonian and a cavity drive term in the control Hamiltonian. The cavity drives were written as real time-dependent fields ($x(t),y(t)$) acting on the quadratures $\hat{x} = a + a^\dagger$ and $\hat{y} = -i (a - a^\dagger)$. We impose amplitude constraints on the optimal control pulses to satisfy the blockade criterion $\epsilon\sqrt{n_0} < \Omega < \chi $, setting a maximum allowed cavity drive amplitude of $2\pi \times (10-15)$ kHz $\approx \Omega/10$. We also explicitly forbid population of the dressed eigenstates at and above the blockaded level to reduce unwanted leakage.

The pulses produced typically had a bandwidth much greater than $\chi$. While this can be decreased by adding bandwidth constraints during the optimal control pulse generation, here we filtered the pulses to a bandwidth of $\pm\chi/2$ about the cavity frequency after generation, with no detriment to the simulated or experimentally measured fidelities. 

The second approach used the simplified Hamiltonian given by SEqn.~(\ref{reduced_Ham}), which includes only the cavity photon number states below the blockade level ($n_0$) and the cavity drive ($\epsilon$). It correctly incorporates the Stark shifts of the cavity levels, but approximates the blockade as perfect, with leakage minimized solely by constraining the cavity drive strength. This simpler optimal control problem resulted in faster pulse convergence and had similar experimental performance for the state preparation sequences of $\ket{1},\ket{2}$ while blockading $\ket{3}$. This improved convergence arises from not needing to manage interference effects to cancel leakage through the blockaded level. It also allowed for the implementation of a qutrit shift gate operation that simultaneously takes $\ket{0} \rightarrow \ket{1}$, $\ket{1} \rightarrow \ket{2}$, and $\ket{2} \rightarrow \ket{0}$.
The pulses heuristically resulted in a gate fidelity of $\sim0.8$, which was $5\%$ worse than the simulated fidelity when including mode and transmon decoherence and decay in accordance with STable~\ref{parameters_table}. The decay and decoherence times of 
 all of the manipulable cavity modes are $\gtrsim 2$ ms~\cite{chakramoriani2020}.

\begin{figure}
  \begin{center}
    \includegraphics[width= 0.5\textwidth]{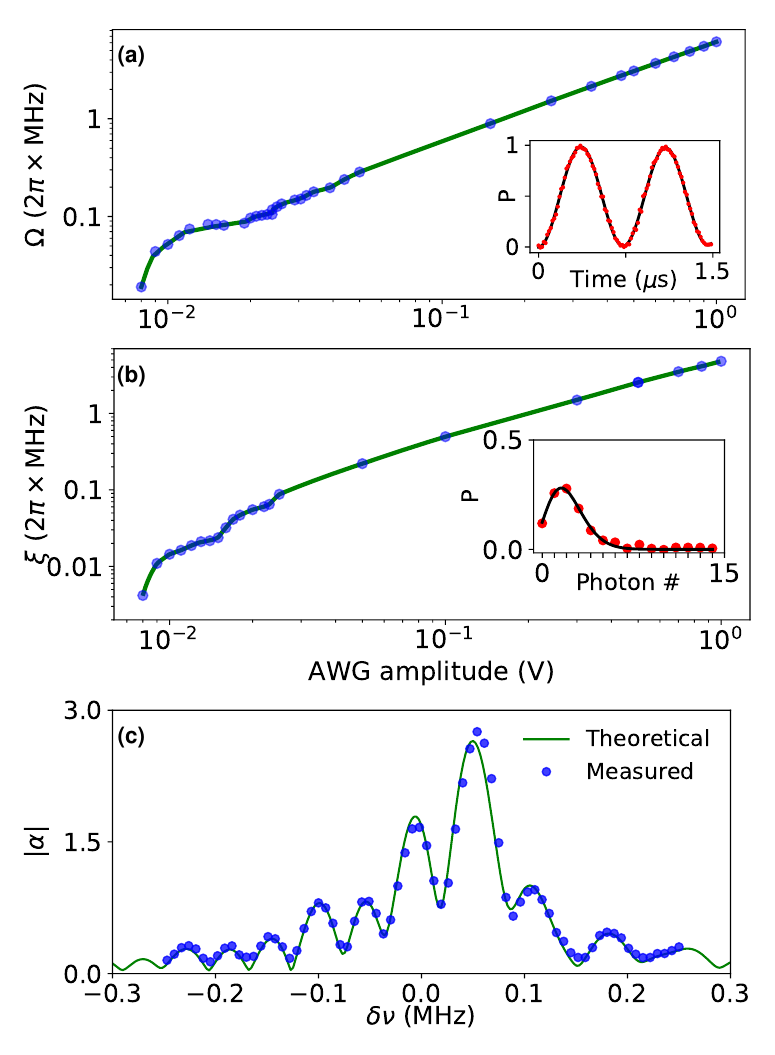}
    \caption{\textbf{Cavity and qubit drive calibrations.} 
    (a) Qubit calibration performed by fitting to Rabi oscillations as a function of AWG amplitude, with an example oscillation shown in the inset. For both the qubit and cavity, the calibration function is not strictly linear at all amplitudes due to digitization effects from the 8-bit control electronics.
    (b) The cavity drive calibration is performed by populating the cavity with coherent states and fitting the resulting photon number Poisson distribution, as a function of AWG amplitude. An example of a prepared state and fit is shown in the inset. The function is calibrated individually for each mode involved in our experiments.
    (c) Fourier transform of the optimal control pulse used to generate Fock state $\ket{1}$ in the cavity. The theoretical and experimental spectrums are in good agreement, particularly around the mode frequency where most of the contribution resides.
    }
  	\label{drive_calibrations}
  \end{center}
\end{figure}

To convert between the optimal control pulse amplitudes (in frequency units) and the control voltages output by the arbitrary waveform generator (AWG), we measured transfer functions for the blockade and cavity drives. The cavity transfer function was determined by driving the target cavity mode for varying times and drive amplitudes and measuring the photon number distribution of the resulting coherent state ($\ket{\alpha}$) using resolved qubit spectroscopy. For a given cavity drive amplitude, we measured $|\alpha|$ as a function of the drive duration ($\tau$) and extract the cavity drive strength from the slope ($|\xi| = 2|\alpha|/\tau$). This process was repeated for different cavity drive amplitudes to obtain the transfer function for the cavity drive strength versus the AWG control amplitude in SFig.~\ref{drive_calibrations}(a). The qubit transfer function was obtained by driving the transmon $\ket{g}\mhyphen\ket{e}$ transition at a fixed amplitude and fitting the resulting Rabi oscillation. The blockade Rabi drive strength $\Omega$ extracted as a function of the control voltage is shown in SFig.~\ref{drive_calibrations}(b). While the transfer functions are linear at higher amplitudes, they become nonlinear at amplitudes $<25$ mV due to  rounding/digitization artifacts from the AWG (8 bit). When input into the experiment, the transfer function data was linearly interpolated with an odd copy reflected about the origin to handle negative drive fields produced by the optimal control. The optimal control pulses are finally shifted back to the lab frame according to $f(t) = x(t)\cos(\omega_{m}t) - y(t)\sin(\omega_{m}t)$, where $\omega_m$ is the frequency of the target mode.

\subsubsection{Measuring the FFT of the optimal control pulses}
The Fourier transform of the optimal control pulse can be measured in-situ by using the cavity as a narrow band ($\sim20$ Hz) spectral filter. We apply the optimal control pulse to the cavity mode (with the blockade drive off) while varying the central carrier frequency, and measure the resulting photon number distribution via resolved qubit spectroscopy. Since the mode only responds on resonance, the magnitude of the resulting coherent state ($|\alpha(\omega_c)|$) as a function of the carrier frequency ($\omega_c$) allows us to determine the FFT of the optimal control pulse. The experimentally measured $|\alpha(\omega_c)|$ and the theoretical FFT of the AWG output pulse with the calibrated transfer function are shown in SFig.~\ref{drive_calibrations}(c). 

\section{Wigner tomography}
\begin{figure}[t]
  \begin{center}
    \includegraphics[width= 0.5\textwidth]{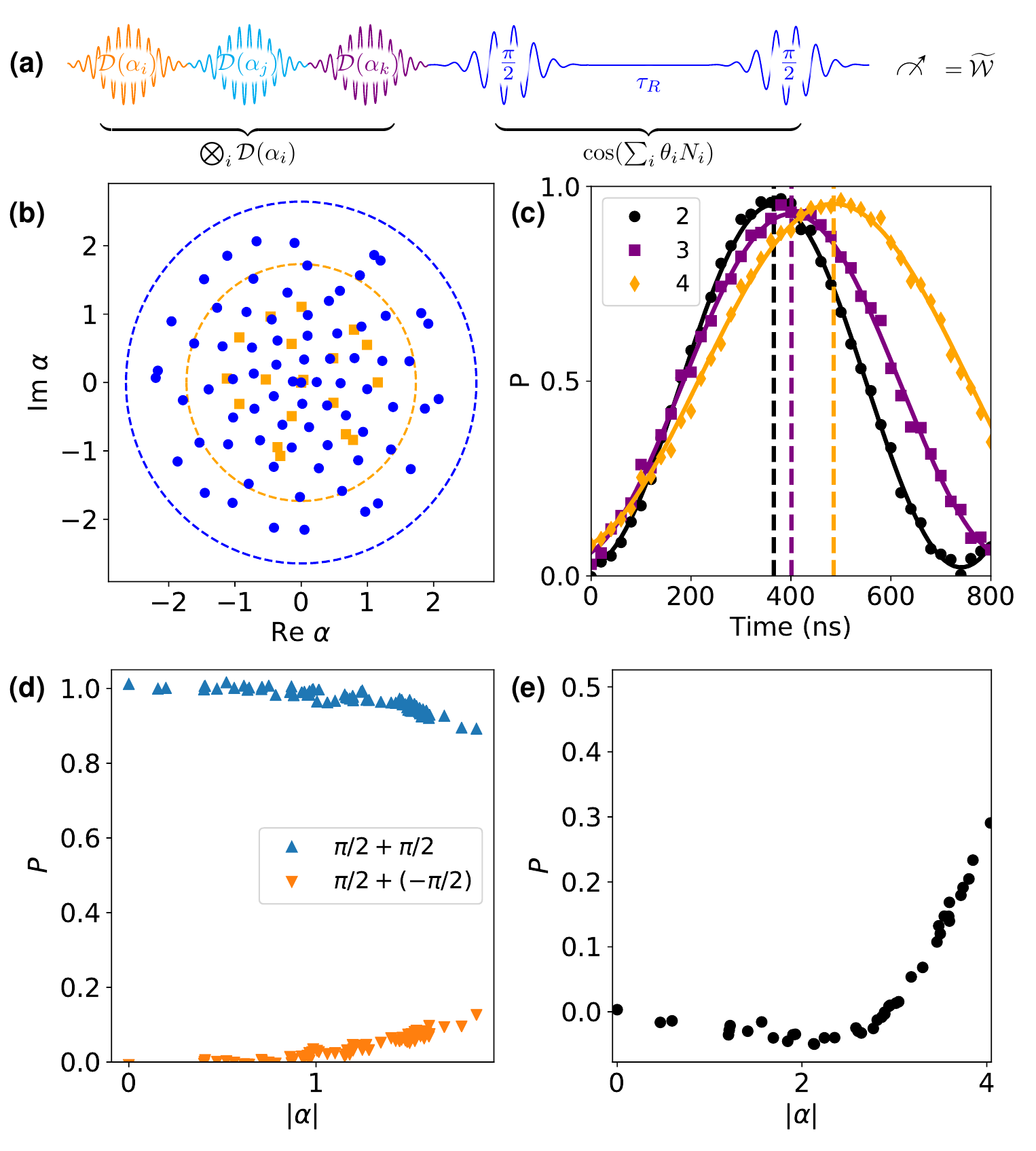}
    \caption{\textbf{Multimode Wigner tomography calibrations.}
    (a) Pulse sequence for Wigner tomography of 3 modes. Sequential cavity displacements are followed by a qubit Ramsey measurement with wait time $\tau_R$.
    (b) Set of cavity displacements used to reconstruct the cavity state with Wigner tomography. The 75 points used in the single mode case (blue) and the 18 points used for each mode in the three-mode tomography case (orange) are shown. Dashed circles indicate the square root of the maximum photon number to which the reconstruction is accurate.
    (c) Ramsey sequence after the addition of a single photon in each mode. Dashed vertical lines indicate the times that correspond to a perfect parity measurement for each mode. This is subsequently used to calibrate the angles $\theta_j$ that correspond to a wait time $\tau_R$.
    (d) Calibrating the bandwidth of the parity measurement. The pulse sequence is a cavity displacement followed by two $\pi$/2 pulses, where the second $\pi$/2 pulse either has phase 0 (blue) or $\pi$ (orange). The finite bandwidth of the qubit $\pi$/2 pulses and the increasing dispersive shift at larger $|\alpha|$ result in imperfect $\pi$/2 pulses, reducing the range of possible measurement results. 
    (e) Calibration of the cross-Kerr between the readout resonator and a cavity mode. The pulse sequence is a cavity displacement of magnitude $|\alpha|$ followed by readout. This does not significantly affect the measurements presented in this work, since the values of $|\alpha|$ used in the Wigner tomography experiments were $<2$. 
    }
  	\label{wigner_tomography_calibrations}
  \end{center}
\end{figure}

Single mode Wigner tomography is performed via a measurement of the photon number parity ($\hat{\Pi}$) following a series of displacements of the cavity mode. This effectively measures the Wigner operator, $\hat{\mathcal{W}}(\alpha) = \hat{\mathcal{D}}_{\alpha}\hat{\Pi}\hat{\mathcal{D}}_{-\alpha}$. The measurements of the Wigner operator for a set of mode displacements $\alpha_i$, $x_{i} = \mathrm{Tr}[\hat{\mathcal{W}}(\alpha_{i})\rho]$ were inverted to reconstruct the density matrix $\rho$ following the procedure described in~\cite{philreinhold}. By converting the Wigner operator and density matrix to vectors, we express $x_i = \langle \langle \mathcal{W}(\alpha_i)| \rho \rangle \rangle $, and construct a matrix $\mathcal{M}$ with $\mathcal{M}_{ij} = \left\langle\langle\mathcal{W}(\alpha_i)\right|_j$ that represents measurements of the Wigner operator at all the displacements. The number of columns of $\mathcal{M}$ is $d^2$, where $d$ is the truncated dimension of the Hilbert space of the cavity up to which the tomography is valid, while the rows correspond to the points in phase space where measurements are sampled ($>d^2$). Since $\mathcal{M}$ is a non-square matrix, we calculate the density matrix by acting the Moore-Penrose pseudoinverse of $\mathcal{M}$ on the vector of measurements $\vec{x}$ , i.e., $\left|\rho\right\rangle\rangle = (\mathcal{M}^{T}\mathcal{M})^{-1}(\mathcal{M}^{T}\vec{x})$. The density matrix extracted from this inversion is made physical by forcing it to have unit trace and imposing positive semi-definiteness. We constrain the trace of the density matrix through the use a Lagrange multiplier ($\lambda$) and perform the inversion as below:
\begin{equation}
\begin{bmatrix}
\left|\rho\right\rangle\rangle\\
\lambda
\end{bmatrix} = 
\begin{bmatrix}
\mathcal{M}^{T}\mathcal{M} & \left\langle\langle I \right|^T \\
\left\langle\langle I \right| & 0 
\end{bmatrix}^{-1}
\begin{bmatrix}
\mathcal{M}^{T}\vec{x}\\
1
\end{bmatrix}.
\end{equation}
Positive semi-definiteness is usually imposed with Cholesky decomposition or other methods. We instead impose that condition using the algorithm presented in~\cite{smolin2014psd}, involving diagonalizing $\mathcal{M}$, iteratively redistributing any negative eigenvalues equally across the remaining positive ones, and subsequently reconstructing $\rho$.

The displacements we perform are to an optimized set of points in phase space that are chosen with the method described in~\cite{philreinhold}, and are shown in SFig.~\ref{wigner_tomography_calibrations}(a). In particular, the set of points minimizes the \textit{condition number} $\kappa$---the absolute value of the ratio of the maximum to the minimum eigenvalue of $\mathcal{M}$. Minimizing $\kappa$ increases the likelihood of $\mathcal{M}$ being invertible, and reduces amplification of error from the inverted $\rho$---measurement noise of magnitude $\epsilon$ results in an error no greater than $\kappa \epsilon$ in the reconstructed density matrix. For our single mode experiments, we used a total of 75 Wigner points with a maximum photon number of 7, with $\kappa = 1.6$. For the multimode measurements of the single-photon W-state, we used 18 Wigner points and a maximum photon number of 3 per cavity mode. 

\subsection{Generalized Wigner tomography} \label{generalized_wt_section}
The parity measurement in Wigner tomography is performed using a qubit Ramsey sequence that is composed of two broadband $\pi/2$ pulses with opposite phases, separated by a wait time $\tau = 1/(2\chi_m)$ during which the qubit acquires a $\theta = \pi$ phase if a single photon is in cavity mode $i$, followed by qubit readout. Since rotations are only distinguishable modulo $2\pi$, all odd (even) photon numbers will place the qubit in the excited (ground) state, resulting in a measurement of photon number parity $\hat{\Pi} = \cos(\pi \hat{N})$. A similar qubit Ramsey sequence that idles for an arbitrary $\tau$ corresponds to a phase shift of $\theta = 2\pi\chi_m\tau$ for a single photon in mode $m$, and a measurement of $\hat{\Theta} = \cos(\theta\hat{N})$. Displacing the cavity mode prior to this general qubit Ramsey sequence allows for the measurement of a generalized Wigner operator $\hat{\mathcal{W}}(\alpha, \theta) = \hat{\mathcal{D}}_{\alpha}\hat{\Theta}\hat{\mathcal{D}}_{-\alpha}$. As long as $\theta$ is known, we can invert the measurements of the expectation value of this operator for a series of known displacements to obtain the density matrix, like in the case of $\theta = \pi$. The error in reconstruction fidelity is dependent on $\theta$, with the smallest error occurring around $\theta = \pi$---where the generalized Wigner function has maximum contrast, and is also the least sensitive to errors in the calibration of $\theta$.

\subsection{Multimode Wigner tomography}
\begin{figure*}[t]
  \begin{center}
    \includegraphics[width= 0.9\textwidth]{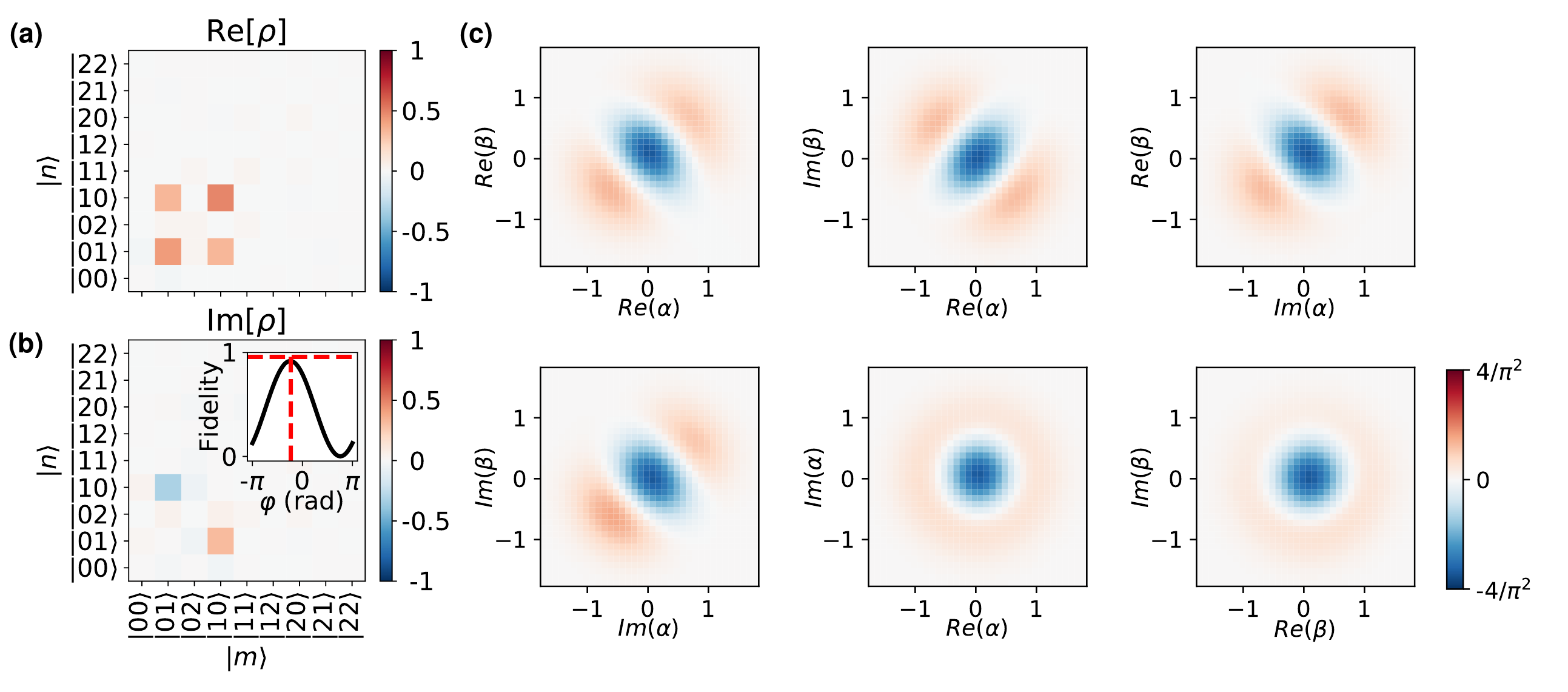}
    \caption{\textbf{Density matrix reconstruction and multimode Wigner tomography data for a two-mode W-state}.
    (a) Real and (b) Imaginary part of the density matrix of the reconstructed two-mode state. The value of $\phi = -0.730$ for the state $(\ket{10} + e^{i\phi}\ket{01})\sqrt{2}$ is determined by the maximum fidelity projection, and is shown in the inset. The corresponding state fidelity is $0.918 \pm 0.012$.
    c) The 6 orthogonal 2D slices of the two-mode Wigner function through the origin of phase space, involving all combinations of the real and imaginary quadratures of both modes. 
    }
  	\label{2mode_tomography_with_angle_inset}
  \end{center}
\end{figure*}

Multimode Wigner tomography has previously been performed via measurements of the joint photon number parity following displacements of each of the cavity modes, effectively corresponding to a measurement of a joint Wigner operator. However, joint photon number parity measurements become challenging when the modes do not have the same $\chi_m$, requiring the use of higher transmon levels~\cite{wang2016schrodinger}, or additional transmons~\cite{ma2020manipulating}. The generalized Wigner tomography protocol described in the previous section provides a workaround, allowing us to replace the joint photon number parity operator $\bigotimes_{m} \hat{\Pi}_m$ with a generalized operator $\bigotimes_m \hat{\Theta}_m$. Since $\theta_m$ need not be identical between modes, a single qubit Ramsey time $\tau$ that corresponds to different $\theta_m$ for each mode $m$ can be utilized to perform the measurement. We can then characterize our state without engineering $\chi_m$. 

In the case of two modes, we used cavity modes 3 and 4 and combined the measurements at three different Ramsey times $\tau_1, \tau_2$, and $\tau_3$ to reconstruct the state density matrix. While only one $\tau$ is necessary, additional times improve the accuracy of the final state reconstruction. The set of $\tau$'s at which we measured was $\{\tau_j \} = \big[419.8, 483.3, 454.0\big]$ ns, which corresponds to $\{\theta_3 \} = \big[\pi, 3.63, 3.39 \big]$ and $\{\theta_4 \} = \big[2.74, \pi, 2.94 \big]$. For these sets of angles, the condition numbers are $\kappa_{3,4} = 1.6$. For the three mode case, we used modes 2,3, and 4 of our cavity, and chose the Ramsey time that corresponds to $\theta_3 = \pi$. We made this choice because $\chi_3$ is between $\chi_2$ and $\chi_4$, resulting in a set of $\left\{\theta\right\}$ that are as close to $\pi$ as possible. This is desirable for reasons as discussed in Section~\ref{generalized_wt_section}. The resulting tomography angles for this single value of $\tau$ are (in radians) $\{\theta_2, \theta_3, \theta_4\} = \big[3.44, \pi, 2.64\big]$. The corresponding values of $\kappa$ are $\kappa_{2,3,4} = 1.6$.

In addition to the density matrices of the multimode W-states presented in the main text, here we provide 2D slices of their multimode Wigner functions. These can be seen for the two and three mode states in SFig.~\ref{2mode_tomography_with_angle_inset} and \ref{3mode_tomography_with_angle_inset}, respectively. The slices correspond to all pairwise combinations of real and imaginary quadratures of each of the modes, leading to 6 slices in the two-mode case and 15 in the three-mode case.

\subsection{Multimode state phase determination and gauge freedom}
\begin{figure*}[t]
  \begin{center}
    \includegraphics[width= 0.9\textwidth]{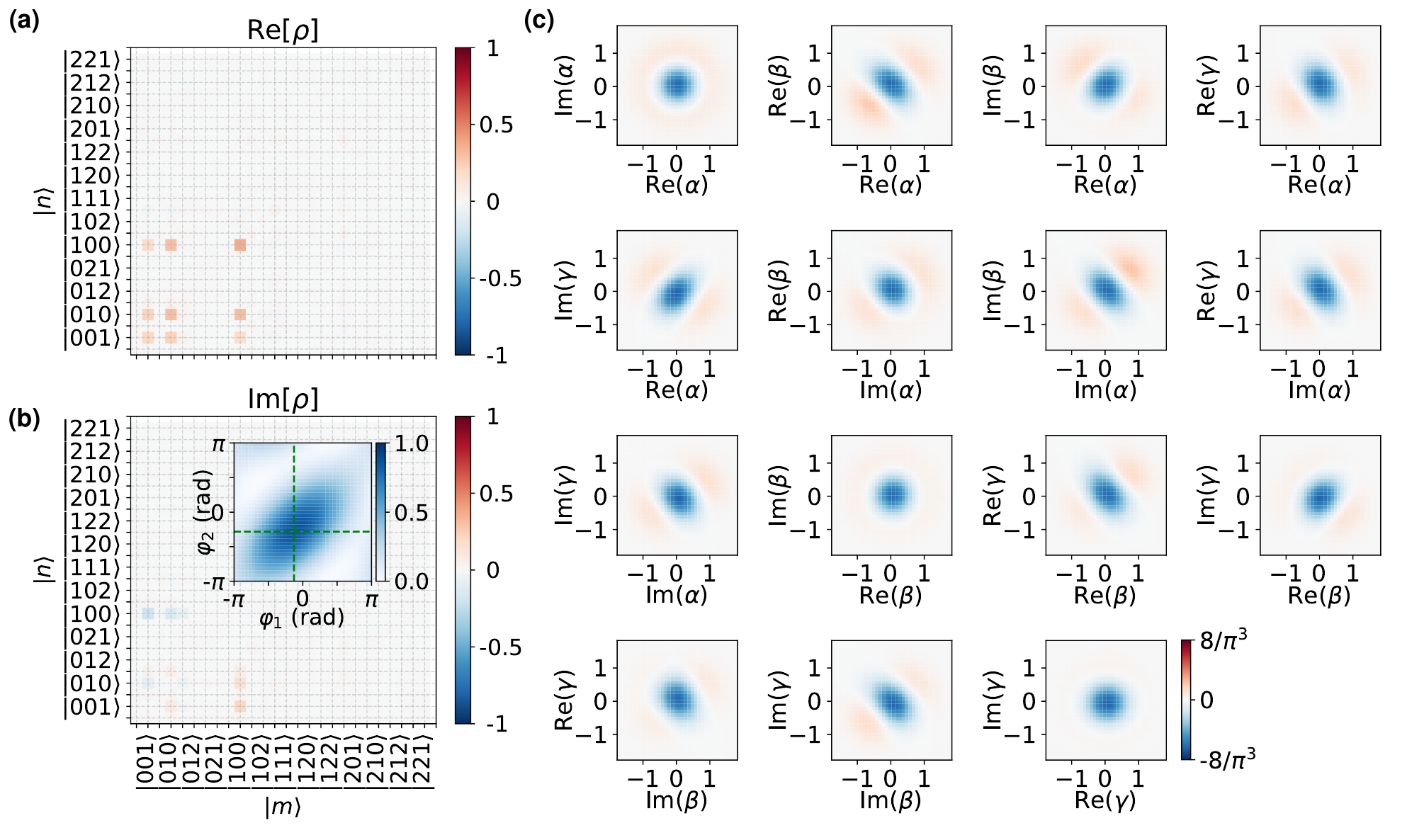}
    \caption{\textbf{Density matrix reconstruction and multimode Wigner tomography data for a three-mode W-state}.
    (a) Real and 
    (b) Imaginary part of the density matrix of the reconstructed three mode state. The values of $\phi_1 = -0.403, \phi_2 = -0.866$ of the prepared state $(\ket{100} + e^{i\phi_1}\ket{010} + e^{i\phi_2}\ket{001}) / \sqrt{3}$ are again determined by the maximum fidelity projection, and  shown in the inset. The corresponding state fidelity is $0.864 \pm 0.014$.
    (c) The 15 orthogonal 2D slices of the three-mode Wigner function through the origin of phase space, involving all combinations of the real and imaginary quadratures of the three modes. 
    }
  	\label{3mode_tomography_with_angle_inset}
  \end{center}
\end{figure*}
We determine the phases $\phi_j$ of our multimode W-states, $(\ket{10} + e^{i\phi}\ket{01})\sqrt{2}$ in the two mode case and $(\ket{100} + e^{i\phi_1}\ket{010} + e^{i\phi_2}\ket{001}) / \sqrt{3}$ in the three mode case, by maximizing the fidelity of the projection onto those states as a function of the $\phi_j$'s. That is, we map our prepared state onto the appropriate (two- or three-mode) expected W-state while sweeping the phase parameters, and pick the angles that give us the projected value closest to 1.
This is shown in the insets of SFig.~\ref{2mode_tomography_with_angle_inset} and SFig.~\ref{3mode_tomography_with_angle_inset}.
We are able to prepare states with different phases by varying the relative phases of our cavity drives. There is a $2\pi$ gauge freedom in the definition of the phase of each cavity mode. For a given choice of these phases, we determine the phase of the prepared states using Wigner tomography. These phases can be  can be modified by a gauge transformation, allowing us to make the reconstructed density matrices real, as in Fig.~5 of the main text.

\subsection{Systematic errors in Wigner tomography} \label{wigner_calibrations_section}
In addition to experimental noise, the Wigner tomography reconstruction has systematic errors that appear in the parity measurement and come from two main sources. The first source is the limited bandwidth of the parity measurement. We mitigate this by using DRAG pulse shaping to maximize the bandwidth of the pulses (Gaussian pulse with $\sigma = 5$ ns). 
The second source is readout error arising from the cross-Kerr interaction between the storage and readout modes. This results in a systematic shift in the readout voltage of transmon states that depends on the number of photons in the storage modes. We calibrate both these errors using the protocols described below, and use them to correct the Wigner tomography. 
The correction to the bandwidth of the parity measurement is calibrated by displacing the cavity to a phase space point used in the Wigner tomography and subsequently applying two $\pi/2$ pulses with either the same or opposite phase, with no wait time in between. This would ideally place the qubit in either the excited or ground state, corresponding to $P_e = $ 1 or 0, respectively. However, despite the large bandwidth of the qubit $\pi/2$ pulses, the dispersive shift takes the pulses off resonance for larger $|\alpha|$. As shown in SFig.~\ref{wigner_tomography_calibrations}(d), this reduces the contrast of the parity measurement and therefore the Wigner operator measurement. We compensate for this effect by scaling the Wigner operator measurement for a given state and displacement using a linear transformation ($\mathcal{W}(\alpha,\rho) \rightarrow a\mathcal{W}(\alpha,\rho) + b $), that takes the upper and lower bounds for the parity measurement ($c_1, c_2$) to their ideal values ($1, 0$), i.e. $a,b = 1/(c_1-c_2), -c_2/(c_1-c_2)$.
This correction is performed for each Wigner point used in the tomography, with varying calibrated values of $a$ and $b$. Wigner points with larger values of $|\alpha|$ deviate more from the ideal 0 to 1 range, as the magnitude of the cross-Kerr and dispersive shift effects scales with |$\alpha$|.

\section{Master equation simulations of blockade dynamics}
We simulate multimode blockade dynamics using a master equation that includes the decay of the cavity modes ($\kappa_m = 1/T^m_1$), as well as the decay ($\gamma_q = 1/T^q_1$) and dephasing ($\gamma^q_\phi =  1/T^q_2 - 1/(2T^q_{1})$) of the transmon:
\begin{equation}
\begin{split}
\dot{\hat{\rho}} = &-i[\hat{H},\hat{\rho}] + \sum_m\kappa_m\mathcal{D}[\hat{a}_m]\hat{\rho} + \gamma_q n^{\mathrm{th}}_{q}\mathcal{D}[\ket{e}\bra{g}]\hat{\rho} \\
&+\gamma_q(1+n^{\mathrm{th}}_{q})\mathcal{D}[\ket{g}\bra{e}]\hat{\rho} + \gamma^\phi_q\mathcal{D}[\ket{e}\bra{e}]\hat{\rho}.\\
\end{split}
\label{master equation}
\end{equation}
Here, $\mathcal{D}$ is the Lindblad dissipator, and $\hat{H}$ is the blockade Hamiltonian given by SEqn.~(\ref{blockade_ham_rf}) for the single-mode case. We include the thermal occupation of the transmon ($n^{\mathrm{th}}_q = 1.2\pm0.5\%$), but ignore the thermal population of the storage cavity modes ($n_m^{\mathrm{th}} \leq 0.01\%$). For the case of multiple cavity modes and drives, this Hamiltonian---written in a frame co-rotating with the blockade ($\nu_b = \nu_q + \delta\nu_b$) and cavity mode frequencies ($\nu_m$), generalizes to:
\bea
\hat{H} &=& \left\{\sum_m\chi_m\hat{a}_m^{\dagger}\hat{a}_m - \delta\nu_b\right\}\ket{e}\bra{e} \nonumber \\ &+&\sum_m\frac{k_m}{2}\hat{a}_m^{\dagger}\hat{a}_m\left(\hat{a}_m^{\dagger}\hat{a}_m-1\right) + \sum_{m\neq n}\frac{k_{mn}}{2}\hat{a}_m^{\dagger}\hat{a}_m\hat{a}_n^{\dagger}\hat{a}_n \nonumber \\
&+& \{\Omega\ket{g}\bra{e} + \sum_m\epsilon_m(t)\hat{a}_m+\mathrm{c.c.}\},
\label{blockade_ham_mm}
\eea
where $\Omega$ is the blockade Rabi frequency, $k_{m}, k_{mn}$ the self and cross-Kerr interactions, and $\epsilon_{m} (t)$ the time dependent cavity drive amplitudes. All the drive tones are sent through the readout port and are directly coupled only to the readout resonator ($\hat{H}_d = \epsilon_r\cos(\omega_d t)(\hat{a}_r + \hat{a}_r^\dagger )$). However, the coupling between the transmon and the modes and their resulting dressing leads to effective transmon and storage mode drives when the readout is driven on resonance with either of them. To lowest order in the dispersive approximation, the resulting transmon and storage mode drives are $\Omega \approx \epsilon_r g_r/(2\Delta_r)$ and $\epsilon_m \approx \epsilon_r g_r g_m/(2\Delta_r\Delta_m)$, respectively.
For generating the optimal control pulses, we treat the drives as being directly on the transmon and the storage modes. This approximation is valid because the detuning between the transmon and the readout ($\Delta_r$) and storage modes ($\Delta_m$) is large compared to the coupling strengths ($g_r, g_m$), which we additionally verify using master equation simulations that include the readout cavity/drive.

\subsection{Single-mode optimal control pulses}
\begin{figure}
  \begin{center}
    \includegraphics[width= 0.5\textwidth]{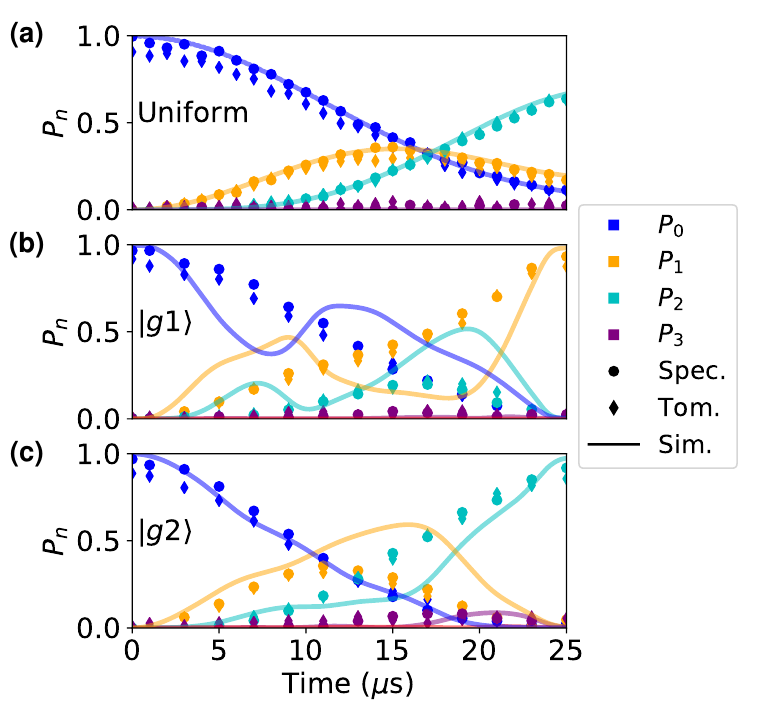}
    \caption{\textbf{Simulations of single-mode optimal control}.
    State evolution vs. time produced by a uniform cavity drive (a) and optimal control pulses and optimal control pulses that prepare $\ket{g1}$ (b) or $\ket{g2}$ (c), in the presence of a blockade drive at $n=3$. Populations were measured directly using qubit spectroscopy (circles), and were also extracted from the density matrices obtained from Wigner tomography (diamonds). The results of the master equation simulations are represented by the solid lines.}
  	\label{optimal_control_sims}
  \end{center}
\end{figure}

\begin{figure}[t]
  \begin{center}
    \includegraphics[width= 0.5\textwidth]{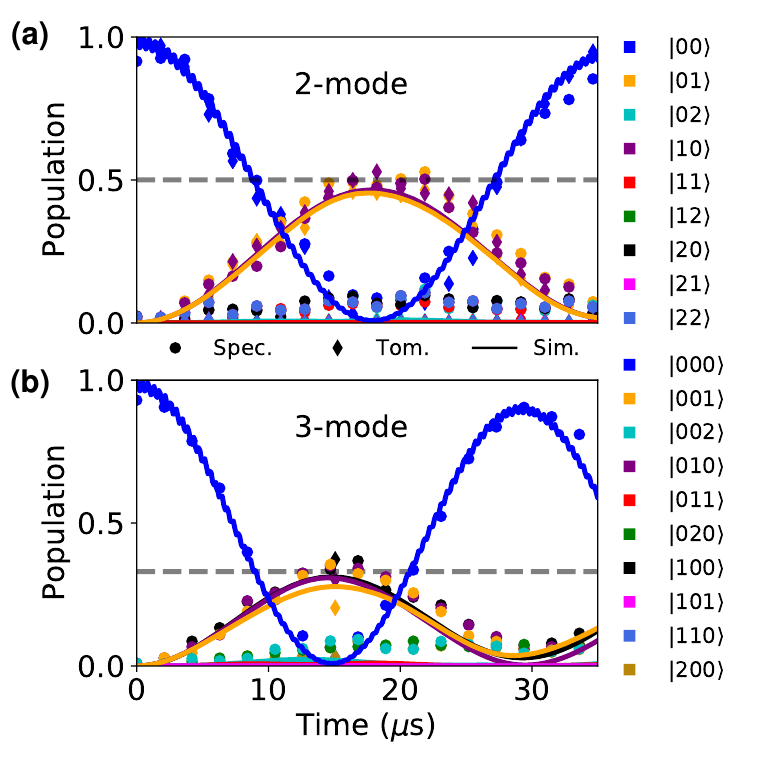}
    \caption{\textbf{Simulations of two- and three-mode W-states}. Populations of different multimode states in the presence of uniform cavity drives and a blockade drive at the mean of the frequencies corresponding to two photons for (a) two ($\nu_3,\nu_4;~ \Omega/(2\pi) = 207$ kHz) and (b) three cavity modes ($\nu_2, \nu_3,\nu_4;~\Omega/(2\pi) = 227$ kHz). For two cavity modes, in addition to monitoring populations through qubit spectroscopy (circles), we extract populations from the reconstructed density matrices obtained from Wigner tomography (diamonds). For the 3 mode case, this comparison to tomography is only made for a pulse corresponding to the preparation of the three-mode W-state ($\tau = 15 \mu s$). In both cases, the cavity drive strength on each mode was $\epsilon/(2\pi) = 10$ kHz. The populations obtained through master equation simulations in the presence of transmon decoherence and cavity decay are represented by the solid lines. 
    } 
  	\label{W-state prep simulations}
  \end{center}
\end{figure}

\begin{figure*}[t]
  \begin{center}
    \includegraphics[width= 0.95\textwidth]{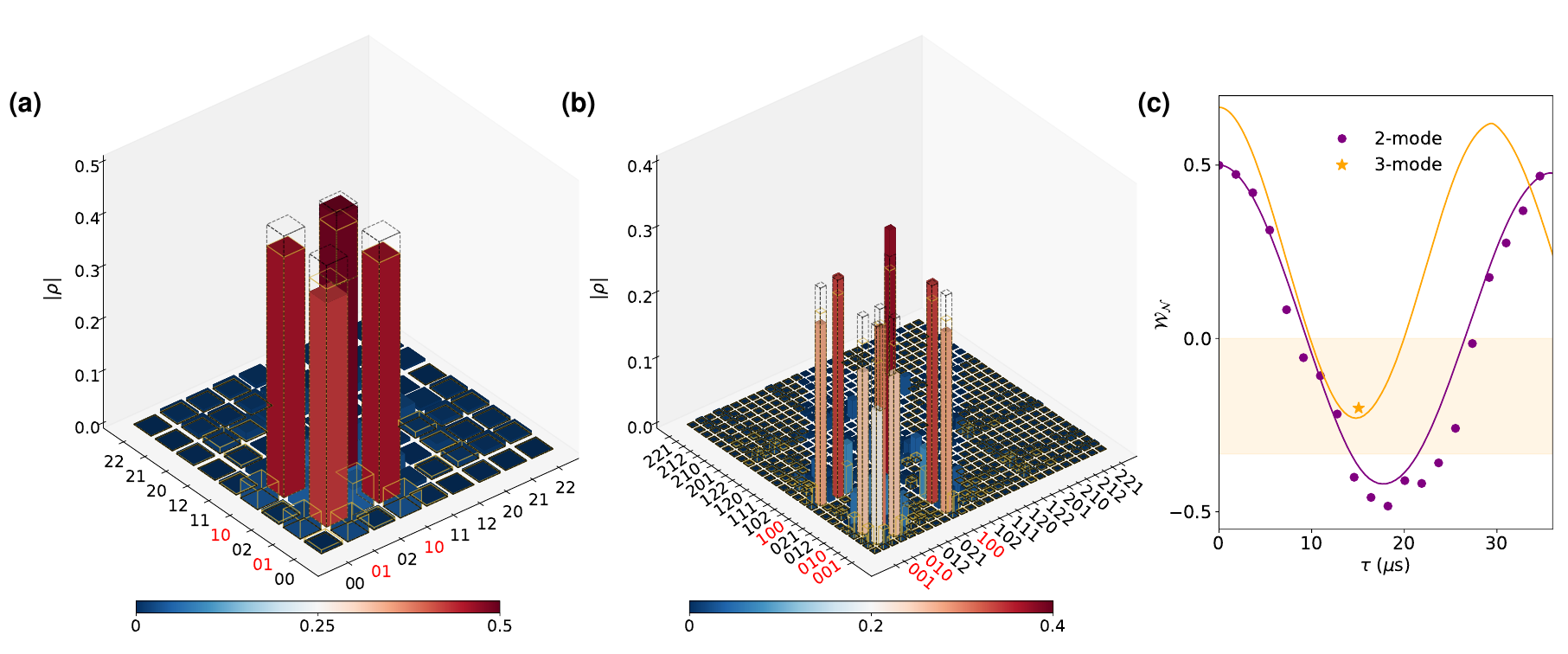}
    \caption{\textbf{Comparison of simulated, prepared, and ideal multimode W-states.} 
    (a) Absolute value of the two-mode W-state density matrix. Populations are represented with colors ranging from red to blue. Dashed black boxes indicate the ideal W-state populations, while yellow boxes show the simulated populations after accounting for the effects of cavity and transmon decoherence and decay.
    (b) Same as (a), but for the three-mode case.
    (c) W-state entanglement witness for 2 and 3 modes. The purple circles correspond to measurements of the two-mode witness as function of the cavity pulse duration in the presence of the blockade drive. The orange $\star$ corresponds to the measured witness for the three-mode state at the time corresponding to the W-state. The solid lines represent the result of master equation simulations of the witness. The orange band between -1/3 and 0 are the witness values that indicate W-state-like tripartite entanglement. 
    } 
  	\label{W-state witness measurement}
  \end{center}
\end{figure*}

We first use the master equation simulations to obtain the expected fidelity in experiments involving the blockade of a single cavity mode, corresponding to the data presented in Fig.~3 of the main text. An analysis of the population evolution generated by a uniform cavity drive, and the optimal control pulses used to prepare $\ket{1}$ and $\ket{2}$ while blockading $\ket{3}$, is shown in SFig.~\ref{optimal_control_sims}. 

The populations are extracted from the density matrices reconstructed from Wigner tomography (diamonds), as well as through number resolved qubit spectroscopy (circles), from the raw data presented in Fig.~3 of the main text. We note that we account for the measurement error arising from the decay during the resolved qubit pulse used for the spectroscopy---a Gaussian pulse with $\sigma = 0.9~\mu$s (duration $= 4\sigma$), by normalizing by the height of the $\ket{0}$ peak obtained from spectroscopy of the vacuum state. No such normalization is performed for the density matrices obtained from the Wigner tomography, which accounts for the slightly lower extracted populations. Apart from this difference ($\sim 5\%$), the populations extracted using both these methods are found to be consistent. 

The master equation simulations of a uniform cavity drive in the presence of a $\ket{g3_i}\leftrightarrow\ket{e3_i}$ blockade drive match well with the experimentally measured populations. We varied the Hamiltonian parameters relevant for the blockade ($\chi,k,\Omega, \delta$) in the simulations, and found that the independently calibrated parameter values also produced the best overlap with the measured state. Despite this, the population trajectories measured by applying slices of the optimal control pulses that prepared $\ket{1}$ and $\ket{2}$ at varying times did not perfectly match with the simulations. Given the close match between the experimental and simulated trajectories for a uniform pulse, the discrepancy in the trajectory is believed to be due to distortions of the optimal control pulse from impedance mismatches along the drive line before reaching the device. While the trajectories themselves deviate, the final states prepared by the optimal control pulses still result in fidelities of $\mathcal{F} = 0.953 \pm 0.022$ ($\ket{1_4}$) and $0.965 \pm 0.022$ ($\ket{2_4}$), compared to simulated fidelities of 0.981 and 0.974, respectively. 

\subsection{Simulations of multimode N-body interactions}
We simulate the dynamics arising from the multimode blockade interactions by using the Hamiltonian in SEqn.~\ref{blockade_ham_mm} in the master equation in SEqn.~\ref{master equation}). The Hamiltonian---co-rotating at the blockade and cavity frequency, is valid for a single blockade drive frequency, as is used in all the experiments presented in this work. 

We prepare two- and three-mode W-states by using a blockade drive detuned from the qubit frequency by the average of the dispersive shifts from adding 2 photons in any combination of modes. For the two-mode case, we study the temporal evolution arising from uniform (and equal strength) cavity drives on both modes (3, 4) in the presence of the blockade drive using photon number resolved qubit spectroscopy and two-mode Wigner tomography. The extracted populations in the different multimode Fock states are shown as a function of the drive duration in SFig.~\ref{W-state prep simulations}(a). At a time $\tau = 18.7~\mu s$, this produces an entangled two-mode W-state, $\ket{\psi} = (\ket{10} + e^{i\phi}\ket{01})\sqrt{2}$. The measured fidelity from Wigner tomography ($\mathcal{F} = \mathrm{Tr}\left[\rho_W\rho\right] = 0.918 \pm 0.012$), and that obtained from master equation simulations ($0.919$) were consistent.  

A similar comparison between the experiment and master equation simulations for the three-mode W-state preparation sequence is shown in SFig.~\ref{W-state prep simulations}(b). Here, the populations are measured as a function of the cavity drive duration using photon number resolved qubit spectroscopy. For both this experiment and the two-mode case, we used a longer resolved qubit $\pi$ pulse (Gaussian pulse with $\sigma \sim 3~\mu s$) than in the single mode case, in order to accurately resolve the differences between the dispersive shifts. 

At the cavity drive duration that corresponds to the W-state, we reconstruct the state using three-mode Wigner tomography, resulting in the density matrices and Wigner functions shown in SFig.~\ref{3mode_tomography_with_angle_inset}. The simulated fidelity of the three-mode W-state was $\mathcal{F} = 0.896$, compared to the experimental measured value of $0.864 \pm 0.014$. A comparison of the simulated, measured, and ideal state populations that produce these fidelities is shown in SFig.~\ref{W-state witness measurement} for both the two- and three-mode W-states. The simulations (yellow edge boxes) include loss and decoherence effects from the transmon and cavity, as well as the qubit temperature. The simulated and measured data are generally in good agreement. The ideal W-states (black dashed edge boxes) are included to serve as a guide. In order for the blockade drive to simultaneously address the dispersively shifted peaks corresponding to 2-photons in any combination of modes, we pick a blockade Rabi strength ($\Omega/(2\pi) = 227$ kHz) which is roughly twice that used in the single-mode blockade experiments, resulting in $~5\%$ higher participation of the transmon in the cavity levels from off-resonant dressing. 

From the reconstructed density matrices for two- and three-mode W-state preparation shown in SFig.~\ref{W-state witness measurement}(a) and (b), we additionally extract the W-state entanglement witness~[S7,S8], 
\begin{equation}
\hat{\mathcal{W}}_{\mathcal{N}} = \frac{N-1}{N} - \ket{W_N}\bra{W_N}
\Rightarrow\left<\hat{\mathcal{W}}_N\right> = \frac{N-1}{N} - \mathcal{F}.
\label{witness}
\end{equation}
where $N$ is the number of entangled modes. For each measured density matrix $\rho$, we extract the witness by sweeping the free phases that characterize the W-state to maximize the state fidelity $\mathcal{F} = \mathrm{Tr}[\rho\rho_{W}]$. The results of these measurements are presented in SFig.~\ref{W-state witness measurement}(c). In the two-mode case, we evaluate the witness as a function of the duration of the cavity drive, while in the three-mode case, we evaluate it at the single time corresponding to the creation of the W-state. We note that $-1/3 < \left<\hat{\mathcal{W}}_{3}\right> = -0.2 < 0 $ indicates genuine tripartite entanglement for the three-mode W-state (the orange band in SFig.~\ref{W-state witness measurement}). 

\bibliography{thebibliography}